\documentclass[10pt,A4paper,amsmath,amssymb,aps,etoolbox,floatfix,
longbibliography,nofootinbib,noshowpacs,onecolumn,preprintnumbers,reprint,
showkeys,showpacs,superscript address]{revtex4}
\newcommand{\bea}{\begin{eqnarray}}
\usepackage[paperheight=11in,paperwidth=9in]{geometry}
\newcommand{\beq}{\begin{equation}}

\newcommand{\eea}{\end{eqnarray}}
\newcommand{\eeq}{\end{equation}}
\usepackage{amsmath}
\usepackage{amsfonts}
\usepackage{amssymb}
\usepackage{bm}       % bold math
\usepackage{comment}
\usepackage[usenames,dvipsnames]{color}
\usepackage{dcolumn}  % Align table columns on decimal point
\usepackage{epsfig}
\usepackage{float}
\usepackage{graphics}
\usepackage{graphicx} % Include figure files
\usepackage{natbib}
\usepackage{hyperref}
\usepackage{capt-of} % to put figure and table in one float, with proper captions
\usepackage{placeins} % for \FloatBarrier
\usepackage{psfrag}
\usepackage{times}
\usepackage[varg]{txfonts}
\usepackage{xcolor}
\usepackage{xspace}

\begin{document}

\title{A critical discussion on the $H_0$ tension}

\author{Salvatore Capozziello}

\affiliation{{\mbox Dipartimento di Fisica "E. Pancini", Universit\`{a} degli Studi di Napoli  "Federico II",}\\ 
{\mbox Complesso Universitario Monte S. Angelo, Via Cinthia 9 Edificio G, 80126 Napoli, Italy,}}

\affiliation{{\mbox Istituto Nazionale di Fisica Nucleare (INFN), Sezione di Napoli,}\\
{\mbox Complesso Universitario Monte S. Angelo, Via Cinthia 9 Edificio G, 80126 Napoli, Italy,}}

\affiliation{{\mbox Scuola Superiore Meridionale, Largo San Marcellino 10, 80138 Napoli, Italy.}}

\author{Giuseppe Sarracino}

\affiliation{{\mbox INAF-Osservatorio Astronomico di Capodimonte,}\\
{\mbox Via Moiariello 16, 80131 Napoli, Italy}}

\author{Giulia De Somma}

\affiliation{{\mbox Istituto Nazionale di Fisica Nucleare (INFN), Sezione di Napoli,}\\
{\mbox Complesso Universitario Monte S. Angelo, Via Cinthia 9 Edificio G, 80126 Napoli, Italy,}}
\affiliation{{\mbox INAF-Osservatorio Astronomico di Capodimonte,}\\
{\mbox Via Moiariello 16, 80131 Napoli, Italy}}
\affiliation{{\mbox INAF-Osservatorio Astronomico d'Abruzzo,}\\
{\mbox Via Maggini sn, 64100 Teramo, Italy}}

\date{\today}
\begin{abstract}
A critical discussion on the $H_0$ Hubble constant  tension is presented  by considering both early and late-type observations. From recent precise  measurements, discrepancies emerge when comparing results for  some cosmological quantities obtained at different redshifts. We highlight the most relevant measurements of $H_0$  and propose potential ideas to solve its tension. These solutions concern  the exploration of new physics beyond the $\Lambda$CDM model or the  evaluation of  $H_0$ by other methods. In particular, we focus on the role of  the look-back time. \\
\\
\textit{ This manuscript is dedicated to the memory of Alexey A. Starobinsky, who recently passed away. He was a distinguished scientist who greatly contributed to the  developments of modern cosmology and theories of gravity with his deep insights and outstanding results.}

\end{abstract}

\keywords{Observational cosmology; Hubble Tension; look-back time.}

\maketitle
\section{Introduction}
The $\Lambda$ Cold Dark Matter ($\Lambda$CDM) model is considered  the cosmological standard, capable of  describing  the observed Universe by fixing only six free parameters \cite{Planck2020}. These are the dark matter density, the baryon density, the observed angular size of the sound horizon at recombination, the scalar spectral index, the curvature fluctuation amplitude, and the reionization optical depth.

Applying  general assumptions on these parameters, it is possible to derive the other cosmological quantities, including the Hubble constant $H_0$ and the other cosmographic parameters  \cite{Weinberg:1972kfs}. The result is a  self-consistent picture of our Universe in good agreement with observations. 

This relatively simple model is able to describe a large part of the Universe history with good precision, from the end of the so-called inflationary era \cite{Guth1981, Starobinsky1979} to the current epoch. According to the $\Lambda$CDM, our Universe is composed by three major constituents: a cosmological constant $\Lambda$, associated with the so-called dark energy, accounting for approximately $68 \%$ of the total  density, a cold (non-relativistic) dark matter component, which should account for  $27 \%$ of the cosmic pie \cite{zyla-etal-2020, Workman-etal-2022, Riess1998, riess_2007, perlmutter-etal-1999, bahcall-etal-1999, spergel-etal-2003, schimd-etal-2006, mcdonald-etal-2006, bamba-etal-2012, joyce-etal-2015, DES2019, Planck2020, Salucci2021}, and lastly, the remaining $5 \%$, composed by baryonic  matter,   stars, galaxies, and all the luminous  structures. The accuracy of this model is remarkable when compared with cosmological observations, such as the accelerated expansion of the Universe \cite{Riess1998} deduced from the observed light curves of Supernovae Type Ia (SNe Ia). Cosmic acceleration can be addressed  by the presence of a cosmological constant  or, in general, by  some unknown form of  dark energy, acting as a negative pressure in the cosmological equations. 

On the other hand, dark matter was initially introduced to account for the virial theorem applied to clusters of galaxies \cite{Zwicky1933}. Subsequent observations revealed that it is also a fundamental component needed to explain the rotation curves of galaxies, which, otherwise, would not be well fitted by Newtonian dynamics  \cite{Babcock1939, Rubin1970} if  only galactic baryonic components are taken into account.

Despite  its overwhelming successes, the $\Lambda$CDM model presents some critical issues that captured the attention of  scientific  community. The most relevant challenges are the nature of dark energy and dark matter, as well as the ongoing tension of the Hubble constant derived by different measurements at different scales. After decades of  precise measurements and tests \cite{Salucci2021, Schnee2011, Mitsou2015, Xenon2017}, no  direct or indirect  evidence of exotic particles constituting cosmic dark  fluids has been found. Consequently, according to the $\Lambda$CDM model, we have no final answer, at fundamental level, on the constituents  of  the observed Universe for  approximately $95 \%$.

A further  issue is related to the $H_0$ tension, that is the  discrepancy between the  late-type measurements of $H_0$ \cite{Riess_2022}, usually linked to the cosmological ladder \cite{Abdalla:2022yfr, Moresco_2022}, and the early-type ones,  associated with measurements of the Cosmic Microwave Background Radiation (CMBR). The most recent results from two prominent collaborations, SH0ES and Planck, report values of $H_0$ as $73.04 \pm 1.04$ km/(s Mpc) at a $68 \%$ confidence level (CL) for the former \cite{Riess_2022} and $67.4 \pm 0.5$ km/(s Mpc) at  $68 \%$ CL for the latter  \cite{Planck2020}. As it stands, there is a $5 \sigma$ tension between these two measurements \cite{Riess_2022}, which, in principle, should provide the same result. Furthermore, this tension extends beyond these two collaborations and involves several late and early-type observations \cite{Abdalla:2022yfr}.

To address these issues, different approaches have been considered like extensions of General Relativity (GR), on which the $\Lambda$CDM model is based. For instance, a particular extension, known as $f(R)$ gravity \cite{Capozziello:2002rd, Nojiri:2010wj, CapozzielloETG, capozziello_2011, Nojiri:2017ncd, DeFelice:2010aj}), has been considered in cosmological applications \cite{Hu:2007nk, Capozziello:2005ku, Oikonomou:2022irx, Rocco, Bajardi} to address different issues related to the $\Lambda$CDM model, like the late-time dark energy \cite{Capozziello:2002rd,bamba-etal-2012}, and the early inflationary behavior  \cite{Starobinsky1979}. The philosophy of these approaches is that, instead of searching for new exotic  ingredients,  gravitational sector should be improved according to the scales. In this perspective, also the $H_0$ tension could be fixed improving geometry \cite{Cai, CapozzielloETG, Benetti_2021, Nojiri:2022ski}. 

Other alternatives imply  Extended Theories of Electromagnetism or the improvement of the  Standard Model of Particles  \cite{spallicci-etal-2021, Spallicci_2022, Sarracino_2022}.  Furthermore, the $H_0$ tension could be also related to some  fundamental quantum concepts, like the Compton Length and the Heisenberg Uncertainty Principle, applied to the cosmological setting \cite{capozziello-benetti-spallicci-2020, spallicci-benetti-capozziello-2022}. 

Finally, under the standard of  "new physics" a large amount of investigations have been pursued both in early and  late Universe  \cite{Bernal_2016, Mortsell_2018, Vagnozzi_2018, Yang_2018, Poulin_2019, Kreisch_2020, Agrawal_2019, Di_Valentino_2019, Pan_2019, Vagnozzi_2020, Visinelli_2019, Knox_2020, Di_Valentino_2020, Di_Valentino_2020b, Di_Valentino_2021f, Smith_2021, Vagnozzi_2021, Nunes_2021, Cyr_Racine_2022, Anchordoqui_2021, Poulin_2021, Alestas_2022, Smith_2022, Reeves_2023, Poulin_2023, Ester1, Ester2, Ratra_1, Ratra_2}.

A recent research line explores the possibility of a "variable $H_0$ constant" i.e. the idea that the measured value of the Hubble constant might depend on the redshift (i.e. the scale)  at which it is measured \cite{Krishnan_2021, Colgain_2021, Krishnan_2022, Colgain_2022, Colgain_2022b, Colgain_2022c, dainotti_2021a, Dainotti_2022a, Dainotti_2023, Schiavone_2022, Schiavone_2023, Malekjani_2023, Vahe, Hu_2023, Jia_2023}. In this context,  the $H_0$ constant can be evaluated by the  look-back time \cite{Capozziello_2023}. The approach consists in determining $H_0$   at any redshift $z$ starting from the  look-back time of the related sources. 

In this paper, a critical discussion on $H_0$ tension  will be presented. 
The outline is the  following.
Section 2 provides a brief summary of  the $\Lambda$CDM model. Section 3 is devoted to   the most prominent measurements related to  $H_0$. In Section 4, we discuss  the look-back time approach to the $H_0$ tension starting from the results in  \cite{Capozziello_2023}. Section 5, a redshift-dependent  $H_0$ is considered. We explore, in particular,  its consequences on cosmological distances. In Section 6, we discuss the results  and draw conclusions.

\section{A Summary of the \texorpdfstring{$\Lambda$CDM model}{}}

The $\Lambda$CDM model is a straightforward byproduct of GR. Assuming the Cosmological Principle,  the Universe is homogeneous and isotropic beyond a certain scale, which is, more or less, over $\sim 120$ Mpc. This assumption is supported by several observations considering large sets of data \cite{Mecke_1994, Yadav_2010, Wiegand_2014, Abdalla:2022yfr}.  Cosmological Principle is implemented by  the Friedman-Lema\^itre-Robertson-Walker (FLRW) metric \cite{Robertson1935}:

\begin{equation}
    ds^2=c^2dt^2-a^2(t) \biggl[ \frac{dr^2}{1-kr^2}+r^2\Omega^2 \biggr],
    \label{FLRW}
\end{equation}
where $\Omega$ represents the angular component of the metric, $a(t)$ is the  scale factor,  and $k$ is the spatial curvature constant. It can be equal to $-1,0,1$ depending on the curvature of the cosmological spatial submanifold.  According to Eq.\eqref{FLRW}, the Einstein field equations can be recast as:

\begin{equation}
    \biggl(\frac{\dot{a}}{a}\biggr)^2+\frac{kc^2}{a^2}=\frac{8\pi G \rho}{3}+\frac{\Lambda c^2}{3},
    \label{first_friedman_equation}
\end{equation}

\begin{equation}
    \frac{\ddot{a}}{a}=-\frac{4 \pi G}{3}\biggl(\rho+\frac{3p}{c^2} \biggr) + \frac{\Lambda c^2}{3},
    \label{second_friedman_equation}
\end{equation}
which are the Friedman equations leading the cosmological expansion \cite{Friedmann1922}. These equations are complemented by the continuity equation and the equation of state defined as:
\begin{equation}
     \dot{\rho}+3\frac{\dot{a}}{a}\biggl(\rho+\frac{p}{c^2}\biggr)=0
    \label{continuity_equation}
\end{equation}
\begin{equation}
     p=w \rho c^2
     \label{equation_of_state}
\end{equation}
Here $G$ is the gravitational constant, $p$ is the pressure, $\rho$  the density, $w$ is the cosmological equation of state parameter, equal to $-1$ for the $\Lambda$CDM model, and $\Lambda$ is the  cosmological constant. These are the equations on which dynamics of $\Lambda$CDM model is based. The scale factor can be written as a function of the redshift as follows \cite{Weinberg:1972kfs}:
\begin{equation}
    a(t)=\frac{a_0}{(1+z)},
    \label{scale_factor}
\end{equation}
where $a_0=1$ is the scale factor normalized at our epoch. This allows us to write the cosmological distances as a function of the redshift \cite{hogg-1999}. It is possible to rewrite Eq. (\ref{first_friedman_equation}) in terms of the cosmological densities as follows\cite{Nemiroff-2008}

\begin{equation}
    \frac{H^2(z)}{H_0^2}=\Omega_{r}(1+z)^{4}+\Omega_{M}(1+z)^{3}+\Omega_{k}(1+z)^{2}+\Omega_{\Lambda},
    \label{Hz}
\end{equation}
where $H(z)=\dot{a}/a$ is the Hubble Parameter, $H_0$ is the Hubble Constant, i.e. the Hubble parameter derived for $z=0$, thus for the Universe at our epoch, $\Omega_{R}$ is the radiation energy density parameter, $\Omega_{M}$ is the matter density parameter,  where both dark and luminous matter  are taken into account, $\Omega_{k}$ is the "density" associated to the curvature, being equal zero for a flat Universe, and $\Omega_{\Lambda}$ is the density associated with the cosmological constant. All these quantities, apart from $H(z)$, are derived at  present epoch.
From this equation, we can define 
\begin{equation} \label{E(z)}
    E(z)=\frac{H(z)}{H_0}=\sqrt{\Omega_r(1+z)^4+\Omega_M(1+z)^3+\Omega_k(1+z)^2+\Omega_{\Lambda}}~.
\end{equation}
This equation allows us to express  cosmological distances as a function of  $E(z)$. It is worth noticing that $E(z)$ depends only on the redshift and the densities of the today  Universe, while it does not depend directly on $H_0$. The  luminosity distance $d_L(z)$,  derived from the intrinsic luminosity and  the photon flux  received by a given  cosmological source, is:
\begin{equation} \label{luminosity distance}
    d_{L}(z)=(1+z)d_{M}(z)~,
\end{equation}
where $d_{M}(z)$ is the transverse comoving distance. 
It is
\begin{equation} \label{comoving flat}
    d_{\rm M}(z)=\frac{c}{H_0} \int_0^z \frac{dz'}{E(z')}\,,
\end{equation}
for a flat Universe with $\Omega_{K}=0$.\\
For an open  Universe with $\Omega_{K}>0$, it is
\begin{equation} \label{comoving open}
    d_{\rm M}(z)=\frac{c}{H_0 \sqrt{\Omega_{K}}} \sinh \biggl(\frac{H_0 \sqrt{\Omega_{K}}}{c} \int_0^z \frac{dz'}{E(z')} \biggr)\,.
\end{equation}
For a closed Universe with $\Omega_{K}<0$, it is 
\begin{equation} \label{comoving closed}
    d_{\rm M}(z)=\frac{c}{H_0 \sqrt{|\Omega_{K}|}} \sin \biggl(\frac{H_0 \sqrt{|\Omega_{K}|}}{c} \int_0^z \frac{dz'}{E(z')} \biggr)\,.
\end{equation}
The luminosity distance is essential in observational cosmology because it can be associated with the  "standard candles". These are astrophysical objects whose intrinsic luminosity can be derived from some intrinsic physical mechanism. Such a mechanism is generally correlated with  quantities that are independent of the source distance, and so can be used to measure  $d_{L}(z)$ intrinsically. Standard candles are a key component of the cosmic distance ladder and play a crucial role in determining  $H_0$.

Another important tool for the estimation  of  cosmic distance ladder is the angular diameter distance, defined as 
\begin{equation} \label{angular diameter distance}
    d_{A}(z)=\frac{d_{M}(z)}{(1+z)}~.
\end{equation}
It is important because it is linked to the  "standard rulers" (i.e., astrophysical objects whose geometrical features can be deduced from their intrinsic physics). We will describe a particular probe employing this definition.
Finally, another very important distance definition is linked to the look-back time, which is the time the photon  takes to reach us from a certain redshift. It is defined as:
\begin{equation} \label{look-back_time}
    T_{lt}(z)=\frac{1}{H_0} \int_0^z \frac{dz'}{(1+z')E(z')}\,.
\end{equation}
It is strictly connected to the light-travel distance, i.e. the path traveled by the photon to reach us from an astrophysical source in the expanding Universe. It is
\begin{equation}
d_{lt}(z)=cT_{lt}(z).
\end{equation}
The last two equations are the starting point for the analysis presented in \cite{Capozziello_2023}, and the novel discussions presented in this work. We have to note here  that, in all the cosmological distance definitions, $H_0$ plays the role of a normalization constant, as it is not directly involved in the integral functions, which, in turn, depend only on the different cosmological components  and the redshift.

\section{The \texorpdfstring  {$H_0$}{} measurements and the tension}

Over the past decades, several methods and astrophysical sources have been employed to measure $H_0$ with high precision  allowing us to obtain, remarkably, very small uncertainties on the measurements  \cite{Abdalla:2022yfr, Salucci2021}. This new era of precision cosmology is also the main reason for the $H_0$ tension, which is at the center of attention of the scientific community. As previously mentioned, we observe a significant $5 \sigma$ tension between the latest SH0ES and Planck collaborations' results \cite{Planck2020, Riess_2022}, but, as we will see, these derivations are actually representative of two entire sets of measurements: the former of the early-type and model-dependent observations, while the latter of the direct late-type and model-independent ones.

\subsection{Late and Early-Type measurements}

Regarding the late-type measurements, many observations, including those by the SH0ES collaboration, are based on the cosmic distance ladder method. In this approach, each step builds upon the previous one through calibration methods, especially in redshift regions where multiple probes are available. This method enables us to reach relatively deep redshift ranges while preserving the precision provided by low-redshift probes, renowned for their accuracy.

The cosmic distance ladder consists of three primary steps \cite{Riess_2021, Riess_2022}. The first  involves precise geometric distance measurements, allowing us to directly calculate the distances of nearby objects. This step is reliable because depends on a straightforward geometrical method and does not require extensive knowledge of the astrophysical probe used for distance measurement. There are three possible anchors for this first rung: Milky Way Cepheid parallaxes, detached-eclipsing binary measurements in the Large Magellanic Cloud \cite{Pietr_2019}, and the water-maser host NGC 4258 \cite{Reid_2019, Riess_2019a}. These three anchors provide approximately $1\%$, $1.2\%$, and $1.5\%$ precision in the calibration of $H_0$, respectively. In recent years, a strong improvement in the Cepheid parallax measurements has been provided by the European Space Agency (ESA) Gaia mission \cite{Gaia_2016, Gaia_DR1_2016, Gaia_2018, Gaia_2021}. Gaia, designed for astrometry, photometry, and spectroscopy, has created the most accurate 3D map of the Milky Way.
The latest release, covering the first 34 months of observations, has provided the largest dataset of Cepheids ever (around 3000 Milky Way Cepheids) allowing us to measure the parallaxes and consequently the distances of Cepheids with unprecedented accuracy \cite{Ripepi_2023}.

The second step involves primary distance indicators, often Classical Cepheids, which exhibit an intrinsic relationship between their luminosities and periods, the so-called Period-luminosity relation (PL) \cite{Marconi_2013, De_Somma_2020, De_Somma_2021}. Although this step is valuable, it may be influenced by potential systematic effects such as the metallicity dependence of the coefficients of the PL relation, which need to be addressed \cite{Marconi2005, Ripepi_2021, De_Somma_2022, Breuval_2022}.

The third step includes probes like SNe Ia, which use primary distance indicators as anchors in regions where both are detected. SNe Ia, with their higher luminosities, can explore relatively high-redshift regions.

Let us discuss in more detail how the SNe Ia are employed as standard candles, given that they are one of the most important components of the cosmological ladder approach \cite{Meng_2015}. The most updated SNe Ia dataset is the Pantheon+ sample \cite{Scolnic_2022}, which was used for the latest $H_0$ measurements \cite{Riess_2022, Brout_2022} and serves as a natural successor to the earlier Pantheon set. \cite{scolnic-2018}. The Pantheon+ sample is composed of 1701 lightcurves taken from 1550 different SNe Ia, covering a redshift range from $z=0.01$ up to $z=2.26$.

In general, the physics behind these astrophysical objects is well-understood. SNe Ia are the byproduct of the explosions of white dwarfs in binary systems exceeding the Chandrasekhar limit due to mass transfer from their companion star. Since this limit is a fundamental constraint for the stability of all white dwarfs, all SNe Ia light curves share similar features. Specifically, the prevailing model, consistent with the majority of SNe Ia observations, is the single-degenerate Chandrasekhar mass explosion \cite{Hillebrandt_2000}. In this model, the white dwarf accretes mass from its less evolved companion star, typically a red giant, which has a significantly lower density, especially in its outer regions.
SNe Ia light curves are well-fitted by the deflagration model shown in \cite{Nomoto_1984}. They are primarily powered by the $\beta$-decay of the radioactive isotope $^{56}$Ni produced during the explosion \cite{Colgate_1969}.

Observationally, this model predicts that SNe Ia light curves typically exhibit an absolute magnitude around M $\simeq -19$\cite{Carroll_2001}. However, both super-luminous \cite{Filippenko_1992a} and sub-luminous \cite{Filippenko_1992_b} SNe Ia have been observed, suggesting the involvement of more complex mechanisms, such as the delayed detonation scenario for super-luminous SNe Ia \cite{Khokhlov_1991}.
Even so, a phenomenological relation has been observed between the peak magnitude of the light curve and the luminosity decline rate in each SN Ia \cite{Phillips1993}, which makes them a proper standardizable candle. Their use as a cosmological probe is based on the following equation 
\begin{equation}
    \mu_{th, SNe Ia}=m-M=5 \log (d_L)+25,
\label{muth}
\end{equation}
where $m$ is the apparent magnitude of the astrophysical object, $M$ is its absolute magnitude, and the luminosity distance is expressed in Mpc. This quantity is confronted with the detected distance modulus $\mu_{obs}$ of the SNe Ia, from which a best-fit of the desired cosmological parameter (as well as of the absolute magnitude $M$) can be performed by employing the following $\chi$-squared function
\begin{equation} \label{eq_chi2_SNe}
    \chi^2_{SNe Ia}= (\mu_{th}-\mu_{obs})^T\times \mathcal{C}_{SNe Ia}^{-1} \times (\mu_{th}-\mu_{obs})~,   
\end{equation}
 where $\mathcal{C}_{SNe Ia}^{-1}$ is the inverse of the covariance matrix \cite{dainotti_2021a}.

It is worth noticing that the absolute magnitude of SNe Ia is treated as a general parameter. This introduces a degeneracy with $H_0$, which can be resolved either by fixing $M$ to a certain value or through calibration processes involving primary distance indicators like Classical Cepheids in the cosmological ladder approach \cite{Scolnic_2022}.

Furthermore,  alternative probes can be employed in the cosmological ladder approach. For instance, the tip of the red giant branch (TRGB) \cite{Salaris1997}, in place of the Classical Cepheids \cite{Freedman_2019}, or type II SNe \cite{de_Jaeger_2020}, the Tully-Fisher relation for galaxies \cite{Kourkchi_2020}, and surface brightness fluctuations \cite{Blakeslee_2021}, as substitutes for SNe Ia in the corresponding cosmological step. The surface brightness fluctuations have also been used as primary distance indicators \cite{Uddin_2023}).

Regarding the results obtained from these methods, a general consensus has emerged around the values derived by the SH0ES collaboration. Indeed, their data have been reanalyzed using different statistical methods, without a significant modification of the final results \cite{Cardona_2017, Camarena_2020, Dhawan_2018, Burns_2018, Follin_2018, Feeney_2018, Abdalla:2022yfr}. Some investigations have also included Cepheids from outside our Galaxy, to address potential biases in measurements linked to specific Cepheid populations \cite{Riess_2019a}. 

An alternative calibration of the SNe Ia, using the TRGB methodology, has produced results that differ from those provided by the SH0ES Collaboration. For example, studies by \cite{Freedman_2019, Freedman_2020, Freedman_2021} derived a value of $H_0=69.6 \pm 1.9$ km/(s Mpc) at a $68\%$ confidence level, which falls between the values from SH0ES and Planck collaborations. However, there are other studies involving the use of TRGB as calibrators for SNe Ia whose results are consistent with the other late-type observations \cite{Jang_2017, Yuan_2019, Kim_2020, Jones_2022, Dhawan_2022}. This has led to ongoing discussions about TRGB-based observations \cite{Uddin_2023, Scolnic_2023}, particularly regarding methodologies to account for potential systematic effects \cite{Yuan_2019, Freedman_2020} and a possible empirical relation between different TRGB observations, similar to what has been derived for Classical Cepheids \cite{Scolnic_2023}.

As mentioned earlier, the cosmological ladder can be used by considering also other probes, such as the surface brightness fluctuations of galaxies, as alternatives to the SNe Ia, or as an intermediate step between Cepheids and SNe Ia \cite{Khetan_2021}. The results remain consistent with the other late-type measurements, even if there are higher errors in the determinations of $H_0$ from these sources \cite{Cantiello_2018, Blakeslee_2021}. A similar approach can also be employed for SNe type II \cite{de_Jaeger_2020, de_Jaeger_2022}, and the Tully-Fisher relation for galaxies \cite{Kourkchi_2020, Schombert_2020, Kourkchi_2022}, obtaining results which remain consistent with the SH0ES collaboration. 

While the cosmological ladder is an intuitive method for determining $H_0$ independently from the cosmological model, it requires very precise knowledge of the astrophysical processes associated with each used probe, especially in the first steps of the ladder. This is because any potential unaccounted-for systematic effect in the first rung could propagate to subsequent ones, as they are calibrated on the preceding. Therefore, using alternating astrophysical objects for the same step becomes a crucial test, as the cross-test between different collaborations using the same probes, to identify and remove possible systematic issues.

The cosmological ladder framework is not the only possible methodology for estimating $H_0$ at late times. An example is the strong lensing time delay estimates, which are independent of cosmological models but do require assumptions about foreground and lens mass distributions  \cite{Abdalla:2022yfr}. Even with this independent alternative method, the results remain consistent with the other late-type approaches \cite{Bonvin_2016, Birrer_2019, Wong_2019, Millon_2020}. It is interesting to mention that, in \cite{Millon_2020}, a decreasing trend in the measure of $H_0$ with the redshift has been noted, in agreement with the previously mentioned "variable $H_0$ constant" \cite{Krishnan_2021, Krishnan_2021b, dainotti_2021a, Capozziello_2023}, which will be the focus of the following analysis.

As we can see, for the late-type measurements, different probes and different methods provide results that are generally in agreement (apart from some exceptions).  We may conclude that the tension is very unlikely to be due to systematic or statistical problems in the data themselves, but rather due to a more intrinsic, physical issue. Indeed, different averages of the late Universe estimates of $H_0$ are in a $4.5-6.3 \sigma$ tension with  values provided by the Planck Collaboration \cite{Verde_2019, Di_Valentino_2021, Abdalla:2022yfr}.

Let us now discuss measurements of $H_0$ based on assumptions and observations related to the early physics of the Universe. In addition to the latest values derived by the Planck Collaboration \cite{Planck2020}, there are other independent measurements, involving  CMBR, all of which consistently yield lower values for $H_0$ if compared to late-type observations \cite{Hinshaw_2013, Aiola_2020, Dutcher_2021}.

Early-time phenomenology can be traced even at low redshift values, with notable examples derived through various probes including the Baryon Acoustic Oscillations (BAO) \cite{Eisenstein2005, Beutler_2011, Blake_2012, du_Mas_des_Bourboux_2020, alam-etal-2021}, Big Bang Nucleosynthesis measurements of the primordial deuterium  \cite{Cooke_2018}, and weak lensing measurements \cite{DES2019}. These probes yield $H_0$ estimates consistent with those of  Planck Collaboration \cite{alam-etal-2021}, and different data reanalyses support these results \cite{Ivanov_2020, d_Amico_2020, Philcox_2020, Ivanov_2021, Ivanov_2022, Philcox_2022, Chen_2022, Ratra_1, Ratra_2}. However, it is essential  remembering that early-type measurements are model-dependent, and work within the $\Lambda$CDM scenario and the Standard Model. Without these assumptions, constraints on $H_0$ and other cosmological parameters are remarkably loosened \cite{Brieden_2021, Abdalla:2022yfr}. Additionally, these measurements provide estimates for all the six free parameters underlying the $\Lambda$CDM model, which, in turn, are used to derive all the other cosmological parameters from the observation of the CMBR peaks.

Let us go into more detail about the BAO.
They  are also utilized independently for cosmological computations \cite{alam-etal-2021} because they consider another kind of cosmological distance, the angular-diameter one. BAO are widely used in  literature to complement various analyses with other probes,  see Refs. \cite{Dainotti_2022e, Spallicci_2022}, and in standalone cosmological computations \cite{alam-etal-2021}.

BAO are density fluctuations  of the visible baryonic matter, caused by acoustic density waves in the early primordial plasma. As such, they are relics of phenomena  occurred in  early times and observed at lower redshift values such as in cluster formations and galaxy distributions. These phenomena are closely related to the acoustic peaks measured from the CMBR \cite{Eisenstein2005} which result from cosmological perturbations  generating sound waves in the relativistic plasma of the early Universe \cite{Bond_1987}. 

In the past decade, BAO-related measurements have significantly improved in precision \cite{alam-etal-2021}, which has proven to be mandatory for modern cosmological applications. This is because the acoustic features in matter correlations are relatively weak and occur at large scales \cite{Eisenstein_1998, Eisenstein2005}. 

Furthermore, these acoustic peaks are associated with different behavior of ordinary and dark matter when they are solicited by perturbations. Ordinary matter expands as a spherical wave, while dark matter remains in place \cite{Bashinsky_2001}. After this event, both dark matter and baryon perturbation start the formation of large-scale structures. Given that, the central perturbation in the dark matter dominates over the baryonic shell, the acoustic feature is manifested as a single spike in the correlation function at approximately 150 Mpc between pairs of galaxies. This scale is typically close to the sound horizon \cite{Bassett_2009}.

It is important to note that, given their nature, behind each BAO-related measurement, there are tens of thousands of observations regarding large structures such as galaxies or clusters of galaxies. The largest spectroscopic survey to date is the Baryon Oscillation Spectroscopic Survey (BOSS \cite{Dawson_2012}), one of the main objectives of the Sloan Digital Sky Survey (SDSS)-III Collaboration \cite{Eisenstein_2011}. Indeed, this collaboration conducted spectroscopy on over 1.5 million galaxies, generating valuable BAO-related data points. This dataset was later complemented by the extended Baryon Oscillation Spectroscopic Survey (eBOSS \cite{Dawson_2016}), which was the cosmological survey within the SDSS-IV \cite{Blanton_2017}.

As previously stated, BAO may be used as cosmological probes  starting from the angular-diameter distance. However, unlike SNe Ia, BAO-related measurements can vary in their definitions. For instance, in the set composed of 16 BAO employed in \cite{Dainotti_2022e, Spallicci_2022, Sarracino_2022}, compiled from \cite{Beutler_2011, Blake_2012, Ross_2015, du_Mas_des_Bourboux_2020, alam-etal-2021}, some of the data offer information about the following quantity

\begin{equation}
        d_V(z)=\biggr[{d_M}^2(z)\frac{cz}{H(z)} \biggr]^\frac{1}{3},
    \label{eq_dilationscale}
\end{equation}
where $d_M$ is the transverse comoving distance defined in Eqs. (\ref{comoving flat},\ref{comoving open},\ref{comoving closed}) and $H(z)$ is defined in Eq. (\ref{Hz}). Also, other cosmological quantities inferred by BAO are the following parameter
\begin{equation}
        A(z)=\frac{100 d_V(z)\sqrt{\Omega_M h^2}}{cz},
        \label{Aparameter}
\end{equation}
where $h=H_0/(100 km \hspace{1ex} s^{-1} Mpc^{-1})$; 
the so-called Hubble distance

\begin{equation}
        d_H(z)=\frac{c}{H(z)},
        \label{dH}
\end{equation}
  and the comoving distance itself. It is important to emphasize that all these definitions are interconnected, but they are not identical. This is a relevant point when constructing the covariance matrix using these heterogeneous measurements.

Additionally, it is worth noting that the majority of the BAO measurements have been rescaled by a factor denoted as $r_d$, that is the distance between the end of inflation and the decoupling of baryons from photons after the recombination epoch. The value of this factor is approximately $150 Mpc$ and it is defined as \cite{alam-etal-2021}.

\begin{equation}
    r_d=\int_{z_d}^{\infty} \frac{c_s(z)}{H(z)} dz,
    \label{r_d_int}
\end{equation}
where $c_s(z)$ is the sound speed, while $z_d$ is the redshift of the drag epoch, which in turn corresponds to the time when  baryons decouple from the photons. This decoupling typically occurs at a redshift of $z \approx 1020$, a value influenced by the physics of the early Universe. This quantity can be approximated using a formula involving cosmological parameters \cite{Aubourg_2015}, that is

\begin{equation}
    r_d=\frac{55.154 \cdot e^{[-72.3(\Omega_{\nu}h^{2}+0.0006)^2]}}{(\Omega_{M}h^{2})^{0.25351}(\Omega_{b}h^{2})^{0.12807}}Mpc,
    \label{eq_rsfiducialtrue}
\end{equation}
where $\Omega_{b}$ is the baryonic density and $\Omega_{\nu}$ is the neutrino density. It is important  noticing that, although they are observed in lower redshift regions, their link to early-Universe physics implies that the cosmological computations derived from them are consistent with the Planck Collaboration results, including those related to $H_0$ \cite{alam-etal-2021}.

\subsection{Overcoming the tension}

The $H_0$ tension is one of the most compelling problems of modern cosmology, and, as such, both observational and theoretical approaches have been explored by the scientific community to address it. From the former point of view, a new independent window has opened by the observations of gravitational waves \cite{Abbott_2017}. Indeed, these detections have already been used as 'standard sirens', to derive new estimates for $H_0$ \cite{Guidorzi_2017, Soares_Santos_2019, Gayathri_2020, Palmese_2023}. 

At present, the precision of these measurements does not allow us to understand if they reduce the tension or if they are more in agreement with either the early or late-type measurements. However, given that we are still in the early stages of the gravitational waves era, a considerable improvement is expected from future observations, especially from the next generations of gravitational wave detectors. This holds great potential as a completely independent, non-electromagnetic method for measuring $H_0$, providing  a new window to solve the tension issue.

Other important measurable quantities for the cosmological studies are the look-back time and the related age of the Universe, which have been used to infer $H_0$ observationally, as it is the case of the ages of the observed astrophysical objects \cite{Jimenez_2019, Bernal_2021, Boylan_Kolchin_2021, Krishnan_2021b, Vagnozzi_2022, Cimatti_2023}.  

Furthermore, interesting observations of high redshift galaxies have been performed by the James Webb Space Telescope. The observed galaxies appear to have unexpectedly high stellar masses which may be in conflict with the age of the Universe as inferred by Planck \cite{Boylan_Kolchin_2023}. It remains unclear whether this potential discrepancy may be attributed to galaxy evolution models or if it has a cosmological origin.

Another possible independent method to infer $H_0$ is based on the $H(z)$ function and its evolution with the redshift, allowing us to extrapolate $H_0$ by requiring $z=0$. This would be a completely model-independent procedure being able to derive this value, independently from other methodologies. A possible approach for this analysis is based on  the so-called cosmic chronometers \cite{Moresco_2012, Moresco_2022}, i.e. the age evolution of  galaxies, as well as on  techniques like Gaussian Process regression \cite{Holsclaw_2010, Holsclaw_2010b}, or cosmography via different polynomials \cite{bamba-etal-2012, Rocco, Bargiacchi_2021}. These approaches have provided estimates of $H_0$ ranging from values consistent with  Planck collaboration  to  values consistent with  late-type estimates \cite{Abdalla:2022yfr}. It is important to note that an extrapolation of $H_0$ at $z=0$ from the $H(z)$ function could be in contrast with assuming a variable $H_0$. We will discuss this point in the next section. 

From an observational standpoint, another potential approach to address the $H_0$ tension involves the extension of  cosmological ladder to higher redshift, in view to bridge the gap between early and late-type measurements. For this aim, one needs to observe astrophysical objects at high $z$ acting as standard candles. An example is represented by the Gamma-Ray Bursts (GRBs), for which different correlations between their intrinsic physical parameters can be observed \cite{Amati_2002, Ghirlanda_2004, Dainotti_2008, Kumar_2015, Dainotti_2020}, allowing us to use them as distance indicators  \cite{Cardone_2009, Cardone_2010, Ester1, Ester2, Amati_2019, Dainotti_2022e, Dainotti_2022f, Bargiacchi_2023, Dainotti_2022c}. 

Other  promising   high-redshift indicators are Quasars \cite{Risaliti_2015}. For these objects, empirical correlations among physical parameters have been found, so, as in the case of GRBs, they could constitute a formidable and populated set of objects to test the Universe at high redshift \cite{Risaliti_2019, Bargiacchi_2022, Bargiacchi_3, Risaliti_2023, Signorini_2023, Dainotti_2023b}.

From a more theoretical point of view, the idea that new physics could be behind the tension is fascinating the scientific community, especially if one considers that this is not the only issue that the $\Lambda$CDM model, and more in general GR itself, presents \cite{Salucci2021, CapozzielloETG}.  Other notable examples  are the nature of  dark energy and dark matter that seems to escape any  probe at fundamental level. A popular approach to address these issues is to consider extensions of  $\Lambda$CDM model and GR. As previously mentioned, $f(R)$ gravity \cite{CapozzielloETG} and other modified theories \cite{Benetti_2021, Salucci2021} have been applied for various cosmological and astrophysical tests \cite{Psaltis_2008,Feola_2020,Schiavone_2022, Schiavone_2023, Dainotti_2022a, Benetti_2021, Salucci2021}.

These modifications  include the possibility of treating the dark energy component as a variable fluid rather than a cosmological constant \cite{Peebles1988, Ratra1988}. This constant can be represented by a scalar field $\phi$ rolling slowly down a flat component of a potential $V(\phi)$ and giving rise to the models known as quintessence \cite{Copeland2006}. In this sense, the Chevallier-Polanski-Linder (CPL) parameterization for the dark energy component \cite{Chevallier_2001, Linder_2003} is one of the most widely studied modifications to the standard scenario.

Other possible approaches   include models where interactions between dark energy and dark matter are taken into account \cite{Piedipalumbo2012, Bonometto2017}. 
The main issue of this framework is the degeneracy existing between different models, which try to address the same problems in completely different ways, thus not allowing us to achieve a natural and definitive extension of the $\Lambda$CDM model and GR \cite{Di_Valentino_2021e, Perivolaropoulos_2022, Sch_neberg_2022}.

New physics may also be linked to modifications of the underlying phenomenology at both the early and late stages of the Universe \cite{Abdalla:2022yfr, Hu_2023}. Examples  for early epochs include: 
\begin{enumerate}
    \item Early Dark Energy, which behaves as a cosmological constant for $z \leq 3000$, and then decades as fast as the radiation density (or even faster) at late times \cite{Karwal_2016, Poulin_2019, Agrawal_2019} via a slow-roll phase transition. While promising, this approach presents problems from both observational and theoretical perspectives \cite{Hill_2020}. Therefore, it has been proposed a modification, called New Early Dark Energy, where instead of a slow phase transition, we have an almost instantaneous one \cite{Niedermann_2021}. This idea is similar to the aforementioned quintessence for late times.
    \item Extra relativistic degrees of freedom at the recombination, parameterized by the number of neutrino species, $N_{eff}$. According to our current understanding, for active massless neutrino families $N_{eff} \sim 3.044$ \cite{Mangano_2005}. This number affects the inferred value of $H_0$. Various models regarding further dark radiation have been proposed \cite{Jacques_2013, Weinberg_2013, Allahverdi_2014, Di_Valentino_2016}.
    \item Modifying the recombination history, by shifting the sound horizon for BAO at recombination. This can be achieved by either varying the early-time expansion history or by modifying the redshift of recombination. Various methods have been proposed to accomplish this, including exotic scenarios in the early Universe \cite{Hart_2017, Jedamzik_2020, Bose_2021}.
\end{enumerate}

The proposed new physics at late times modifies our interpretation of dark energy in various ways. Apart from what we have mentioned before, we recall the following:
\begin{enumerate}
    \item Considering a cosmological bulk viscous fluid, characterized by a peculiar form of its pressure term, which is made up of two parts, where the first one is the usual component linked to the density via an equation of state, while the second is linked to the  viscosity \cite{Brevik_2005, Wang_2017, Yang_2019b, da_Silva_2021}. 
    \item A chameleon field for dark energy, whose mass varies in accordance with the matter density of the considered environment, and whose variability would imply a measurement on $H_0$ dependent on the particular region in which has been performed \cite{Khoury_2004, Khoury_2004b, Vagnozzi_2021b, Cai_2021, Benisty_2022}.
    \item Diffusion models, implying an interaction between dark energy and dark matter via a non-conserved energy tensor $T_{\mu \nu} $ \cite{Haba_2016, Koutsoumbas_2018, Calogero_2013}, which seems to reduce the $H_0$ tension with different types of matter fields \cite{Perez_2021}.
    \item General dynamical behavior for dark energy, following a philosophy similar to the CPL parameterization. In this sense, one could define emergent dark energy, which has had no effects in the early stages of the Universe, as it completely emerges at late times \cite{Li_2019, Pan_2020, Rezaei_2020}.
    \item A Running Vacuum model, linked to possibly Quantum Field Theory or String Theory, could be used to explain theoretically a possible phenomenological dependence of cosmological and gravitational constants with the cosmic time \cite{Sol__2013, Sol__2015}. This kind of models can actually encompass different assumptions regarding the behavior of dark energy. It has also been successfully tested \cite{Sol_Peracaula_2018, G_mez_Valent_2018}.
    \item The presence of local inhomogeneities that could affect the late-time measurements of $H_0$, which may be either due to possible observational issues like incomplete sky sampling, astrophysical problems like incorrect modelling of the local structures, or a more fundamental nature, like the departure of the FLRW assumption at very small scales \cite{Camarena_2018, Bengaly_2019, Krishnan_2022_b, Kalbouneh_2023,Giani_2024} 
\end{enumerate}

Alternatives to extensions of GR have also been sought. These alternatives include considerations for potential extensions of the Maxwellian Electromagnetism \cite{helayelneto-spallicci-2019} in a cosmological setting, introducing a second, optical component to the measured redshift of astrophysical sources in cosmological models without dark energy \cite{spallicci-etal-2021, Spallicci_2022, Sarracino_2022}. 

In order to represent a valid alternative to the $\Lambda$CDM model, evidence of this kind of effects has to be found. A possible  approach is to investigate upper limits on the  photon mass \cite{bonetti-dossantosfilho-helayelneto-spallicci-2017, bonetti-dossantosfilho-helayelneto-spallicci-2018, scharffgoldhaber-nieto-2010, retino-spallicci-vaivads-2016}, especially considering that some extensions propose  massive  photons \cite{debroglie-1936, proca-1936b}.

Another investigated possibility is dealing with the $H_0$ tension as an evidence of a more fundamental limit on  observations, linked to Quantum Mechanical concepts like the Compton Length and the Heisenberg Uncertainty Principle \cite{capozziello-benetti-spallicci-2020, spallicci-benetti-capozziello-2022}. This approach seeks for addressing the tension without attributing it to unlikely experimental errors or unknown novel physics.
 
The previously mentioned alternatives offer promising solutions for addressing the $H_0$ tension and providing a satisfactory explanation for its existence. The main issue is to determine if one of these alternatives truly resolves the tension or if the solution lies within a completely different framework. To accomplish this issue, one not only needs to provide definitive evidence of peculiar effects beyond the $\Lambda$CDM model, but also to build up an appropriate extension of the standard framework containing not only novel ideas to solve the $H_0$  and other tensions but also retaining the outstanding success achieved by GR and $\Lambda$CDM model when they are compared with observations. 

\begin{figure}
\includegraphics[width=0.8\hsize,height=1\textwidth]{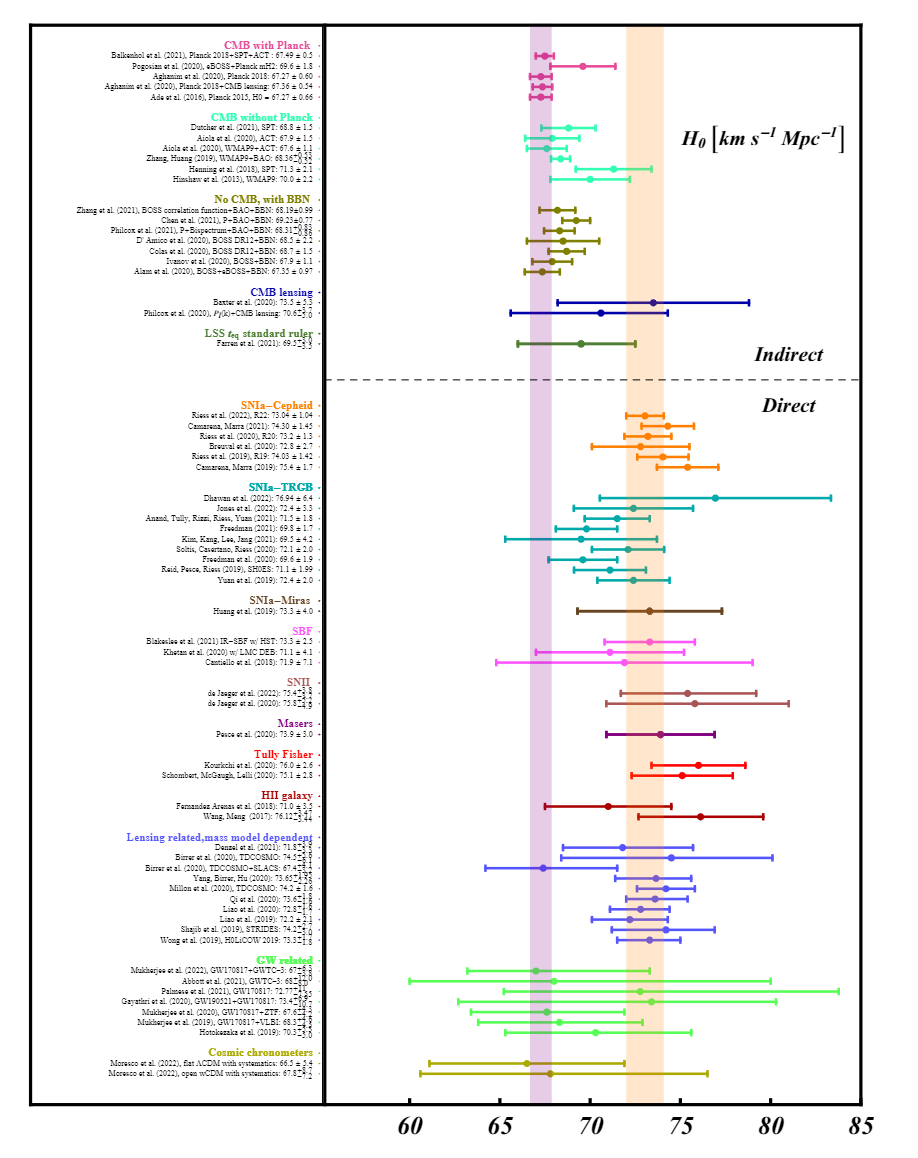}
\caption{A summary of $H_0$ measurements performed both at early and late times. Credits to \cite{Abdalla:2022yfr}.}
\label{Figure_1}
\end{figure}

As said, a novel approach considers the possibility of an evolution of  $H_0$ constant with  redshift as a way to address the tension \cite{Krishnan_2021, Colgain_2021, Krishnan_2022, Colgain_2022, Colgain_2022b, Colgain_2022c, dainotti_2021a, Dainotti_2022a, Dainotti_2023, Schiavone_2022, Schiavone_2023, Malekjani_2023, Vahe, Hu_2023}. In this sense, the evolution of $H_0$ and the corresponding tension are considered as a sort of "diagnostics" of a symptom of the breaking-down of the $\Lambda$CDM model (or even of the FLRW metric, because it would suggest, according to this interpretation, that the parameterization $H(z)=H_0 E(z)$ is not working), marking the points in which observations are not consistent with the model \cite{Krishnan_2021, Colgain_2021, Krishnan_2022, Colgain_2022, Colgain_2022b, Colgain_2022c}. 

In this  context, a dependence of $H_0$ on the redshift has been observed in real SNe Ia data and interpreted through the $f(R)$ gravity formalism \cite{dainotti_2021a, Dainotti_2022a, Dainotti_2023, Hu_2023, Schiavone_2022, Schiavone_2023}.  More specifically, in \cite{dainotti_2021a}, a  functional form 
\begin{equation}
H_0^z=H_0/(1+z)^{\alpha}\,,
\end{equation} 
has been assumed and fitted with the Pantheon set of SNe Ia \cite{scolnic-2018}.  Results show that $\alpha$ is not consistent with 0 within $1 \sigma$, hinting at a smooth, slow, but continuous, decrease of  $H_0$ value with the redshift. A variable $H_0$  could  be due to a possible break-down, at some scale,  of the Cosmological Principle on which the $\Lambda$CDM model is based (for general reviews, see \cite{Abdalla:2022yfr, Perivolaropoulos_2022, Kumar_Aluri_2023}, for discussions  on  diagnostics,  see \cite{Krishnan_2021b, Krishnan_2022, Krishnan_2022_b, Krishnan_2023}. 

It is worth emphasizing that  $H_0$ measurements  discussed in this section represent only a part of those obtained in recent years. For a more comprehensive overview,  refer to Figure \ref{Figure_1} taken from \cite{Abdalla:2022yfr}.

\section{The \texorpdfstring{$H_0$}{} tension and the look-back time}

\subsection{The \texorpdfstring{$T(z)$}{} parameterization}

Let us now delve into the approach proposed in \cite{Capozziello_2023}, where  $H_0$ is derived from the look-back time defined in Eq. (\ref{look-back_time}). We will further discuss and improve this  parameterization. Let us start from showing that, operatively, we can take into account   a Taylor expansion of the scale factor as follows
\begin{equation} \label{expansion}
    a(t)=a_0+ \frac{d(a(t_0)}{dt}[T(z)-T_0]+...=a_0+H_0[T(z)-T_0]+...~.
\end{equation}
We can assume that $a_0=1$ and approximate $H_0=1/T_0$, where $T_0$ is the today Universe age. This  approximation is at $5\%$ if compared with the Planck  data.  It is easy to obtain the integral
\begin{equation} \label{H0_Planck}
      H_0^{(\infty)} =\frac{1}{T_0} \int_0^\infty \frac{dz'}{(1+z')E(z')}~.
\end{equation}
This approximation is mandatory in view to justify our approach. Clearly, considering only the first-order term, we recover the parameterization
\begin{equation} \label{T(z)}
    T(z)=\frac{T_0}{(1+z)}=a(t)T_0~,
\end{equation}
that allows us to label the universe age at various redshifts $T(z)$ starting from  $T_0$. The definition of  look-back time can be recast as:
\begin{equation} \label{H0_look-back}
    H_0=\frac{1}{T_{lt}(z)} \int_0^z \frac{dz'}{(1+z')E(z')}~.
\end{equation}
According to this equation, we can infer $H_0$ from $T_{lt}(z)$ at any $z$. This result is consistent with the age of the Universe, by considering the general definition
\begin{equation}
    T_{lt}(z)=T_0-T(z)~,
\end{equation}
where $T_0$ and $T(z)$ are the Universe age, today and at a given redshift respectively. Considering  Eq. (\ref{T(z)}),  we can derive
\begin{equation} \label{H0_Riess}
    H_0^{(z)}=\frac{(1+z)}{T_0 z} \int_0^z \frac{dz'}{(1+z')E(z')}\,.
\end{equation}
 Eq. (\ref{T(z)}) can be confronted  with other parameterizations.  The idea is  linking  different ages of the Universe with  today epoch, avoiding the  integral  time evolution, depending on  $E(z)$,   as in Eq. (\ref{H0_Riess}). In other words,  one can adopt a point-by-point labeling process. The most natural label that we may use is the scaling factor itself $a(t)$, which expresses how the size of the Universe changes with its expansion. In this perspective, $T(z)$ is a projection of $T_0$ at a given redshift. This is the main reason why we have operatively computed this labeling from Eq. (\ref{expansion}).

 As  noted in \cite{Capozziello_2023}, for $z \to +\infty$, it is
$
    {\displaystyle \lim_{z \to +\infty} \frac{z+1}{z} = 1,}
$
and thus it is easy to recover Eq. (\ref{H0_Planck}) from our approach,
which can also be interpreted as the definition of  Universe age, denoted as $T_0$. This means that the parameterization is in agreement with the  age definition at high values of $z$. Additionally, in Ref. \cite{Capozziello_2023},  it is demonstrated that this parameterization is remarkably consistent with different $H_0$ measurements, ranging from late and early epochs, where  different probes  are taken into account. More specifically, we report here  results obtained  by the Planck collaboration for the following quantities \cite{Planck2020}:
\begin{equation}
    T_0=13.797 \text{Gyr}, \quad \Omega_r=9.252 \times 10^{-5},\quad  \Omega_M=0.3153\,, \quad \Omega_{\Lambda}=0.6847\,.
\end{equation}
which have been compared with the following $H_0$ observations at $68 \%$ CL:
\begin{itemize}
\item $H_0=73.04 \pm 1.04\, \text{km/(s Mpc)}$ from the SH0ES collaboration, inferred by the cosmic distance ladder method considering Classical Cepheids + SNe Ia) up to $z=0.15$ \cite{Riess_2022};

\item $H_0=67.4 \pm 0.5\, \text{km/(s Mpc)}$ from the Planck collaboration, obtained by the CMBR observations at $z \sim 1100$ \cite{Planck2020};

\item $H_0=69.9 \pm 1.9\, \text{km/(s Mpc)}$, obtained by using the TRGB as an anchor for SNe Ia instead of the Classical Cepheids, at $z=0.08$ \cite{Freedman_2019};

\item $H_0=75.8 \pm 5.0\, \text{km/(s Mpc)}$, derived from SNe Type II as the last step of the cosmological ladder, at $z=0.45$ \cite{de_Jaeger_2020}; 

\item $H_0=73.3 \pm 4.0\, \text{km/(s Mpc)}$, derived from the Mira Variables employed as anchors of SNe Ia,  at $z \sim 0.15$ \cite{Huang_2020}; 

\item $H_0=76.0 \pm 2.6\, \text{km/(s Mpc)}$, derived from the Tully-Fisher relation for spiral galaxies, at $z=0.5$ \cite{Kourkchi_2020}; 

\item $H_0=73.3 \pm 2.5\, \text{km/(s Mpc)}$, derived from the surface brightness fluctuations for the galaxies, at $z=0.33$ \cite{Blakeslee_2021}; 

\item $H_0=69.5 \pm 3.3\, \text{km/(s Mpc)}$, inferred from the Large Scale Structure $t_{eq}$ standard ruler, and thus confronted to our computations at the redshift of equivalence $z_{eq} \sim 3300$ \cite{Farren_2022}; 

\item $H_0=72.0 \pm 1.9\, \text{km/(s Mpc)}$, inferred from the masers + SNe Ia and compared at $z \sim 0.15$ \cite{Reid_2019}; 

\item $H_0=73.3 \pm 1.8\, \text{km/(s Mpc)}$, derived from gravitational lensed quasars, confronted at $z=0.745$ \cite{Wong_2019}; 

\item $H_0=67.9 \pm 1.5\, \text{km/(s Mpc)}$, which is a measurement provided by the CMBR independently from the Planck collaboration, and as such corresponding at the reionization epoch $z \sim 1100$ \cite{Aiola_2020}; 

\item $H_0=69.6 \pm 2.1\, \text{km/(s Mpc)}$, linked to the 21 cm absorption line and corresponding at the beginning of the so-called Cosmic Dawn, i. e. when the first stars formed ($z \sim 17.2$), in combination with CMBR data and considering a Chaplygin gas model for the dark sector
\cite{Yang_2019}; 

\item $H_0=73.4\pm 8.8\, \text{km/(s Mpc)}$, deduced by gravitational waves, at $z=0.438$ \cite{Gayathri_2020}.

\end{itemize}
We  remind that the corresponding redshift for each measurement has been determined by considering either the upper limit of the redshift range of the sample used to infer $H_0$ or the redshift associated with the specific physical process.
\begin{figure}
\includegraphics[width=0.45\hsize,height=0.5\textwidth]{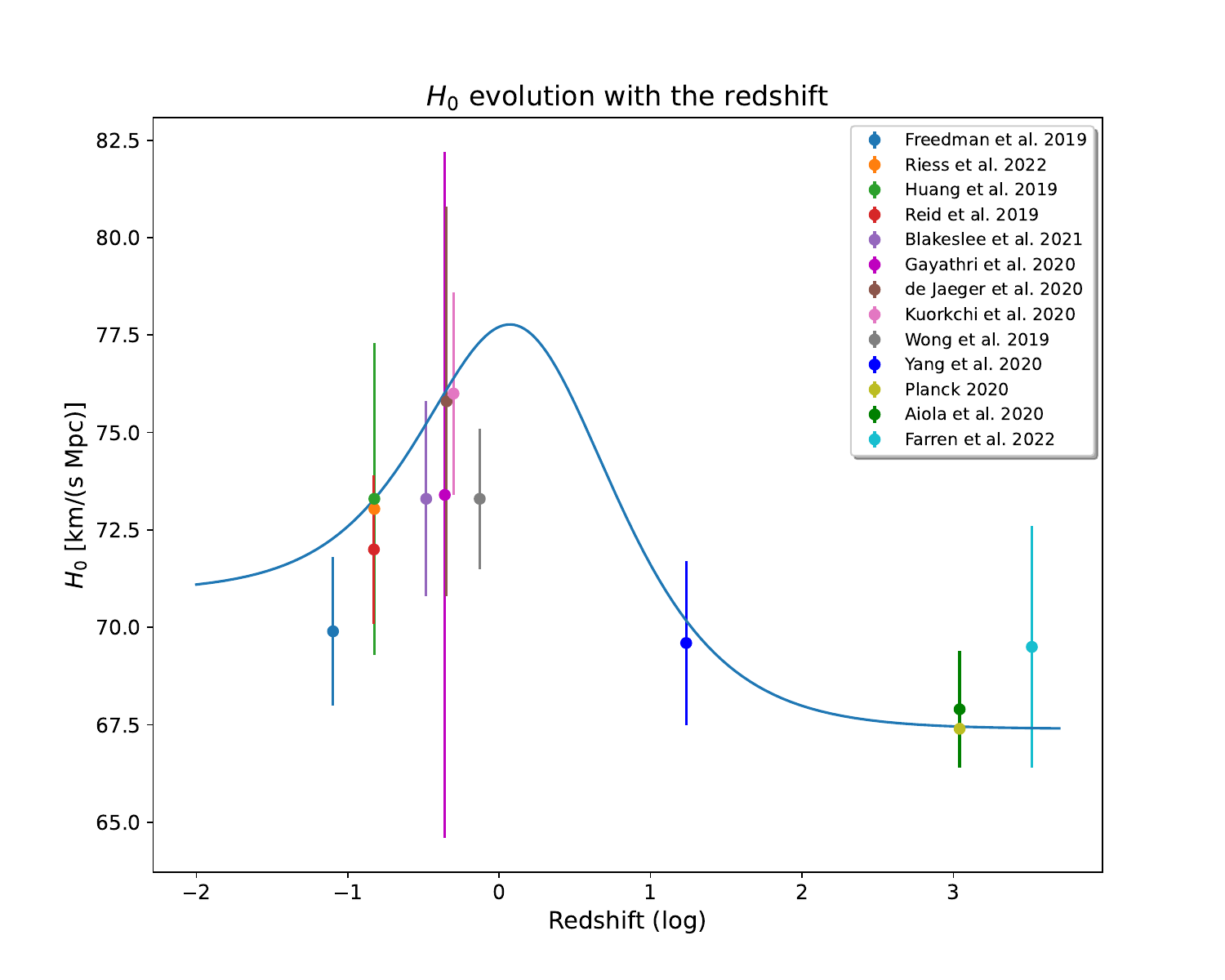}
\includegraphics[width=0.45\hsize,height=0.5\textwidth]{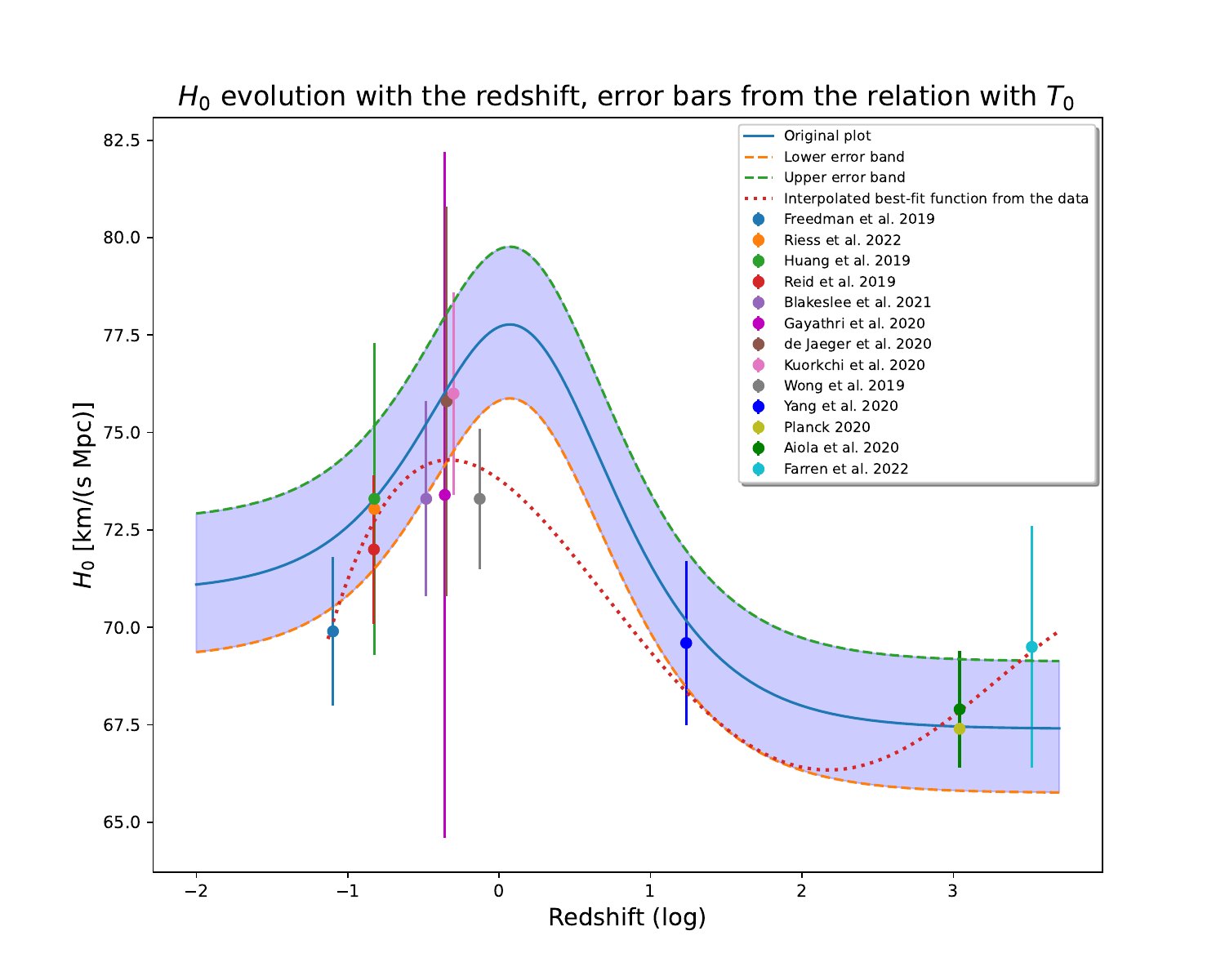}
\caption{Left panel: The values of $H_0$ derived from Eq.(\ref{H0_Riess}) are plotted against the redshift and confronted with observational data  \cite{Capozziello_2023}. For each measurement, the redshift value has been selected either at the upper limit of the redshift range of the sample or at the redshift corresponding to the specific physical phenomenon considered for the estimation (cosmic dawn, recombination, equivalence epoch, and so on). Right panel: the same plots, but considering a $5 \%$ uncertainty on our model linked to the relation between $T_0$ and $H_0$, and also a polynomial fit on the $H_0$ measurements. The $x$-axis is reported in a logarithmic scale. We recall that the measurements have been taken from \cite{Riess_2022, Planck2020, Freedman_2019, de_Jaeger_2020, Huang_2020, Kourkchi_2020, Blakeslee_2021, Farren_2022, Reid_2019, Wong_2019, Aiola_2020, Yang_2019, Gayathri_2020}}
\label{Figure_2}
\end{figure}
Results reported in \cite{Capozziello_2023} are
 displayed in left panel of Fig. \ref{Figure_2}. They will be used   as  reference for our tests. Here, we  consider  also the effects of  $5\%$ approximation in assuming  $H_0=1/T_0$. Essentially,   error bands at $5\%$ can be taken into account. Results are shown in the right panel of Fig. \ref{Figure_2}, where we see that the model is still consistent with the $H_0$ measurements. We have also performed a polynomial fit of the $H_0$ measurements, independent of the cosmological model. It is worth  noticing that  it is consistent with   late and early measurements, while it results shifted  with respect to  the peak. This feature  has to be expected due to  the lack of direct $H_0$ in the intermediate redshift range.

It is important to discuss the value of $T_0$  used in the analysis. This quantity is linked to $H_0$. More specifically, an anti-correlation exists between $T_0$ and $H_0$ has been reported in Ref. \cite{De_Bernardis_2008}. An independent way to test  $H_0$ is to compare measurements with  estimate ages of  old objects  such as stars and globular clusters. 

In this sense, different measurements are consistent with  $T_0$ derived from  early-type observations \cite{Vandenberg_1996, Soderblom_2010, Catelan_2017, Jimenez_2019, Valcin_2020}. This suggests that the $H_0$ value, inferred from late-type observations, would imply a Universe that is too young according to these measurements. Therefore, novel physics might be required to reconcile  $H_0$   with the late-type derivations.

This is the first reason why we  started from the values provided by the Planck Collaboration. The second one is that both $T_0$ and $H_0$, according to  Planck, are derived quantities which have been computed by the six aforementioned free parameters. Therefore, even if $T_0$ and $H_0$ are linked, we do not incur in a circularity problem  \cite{Planck2020}.

\begin{figure}
\includegraphics[width=0.8\hsize,height=0.6\textwidth]{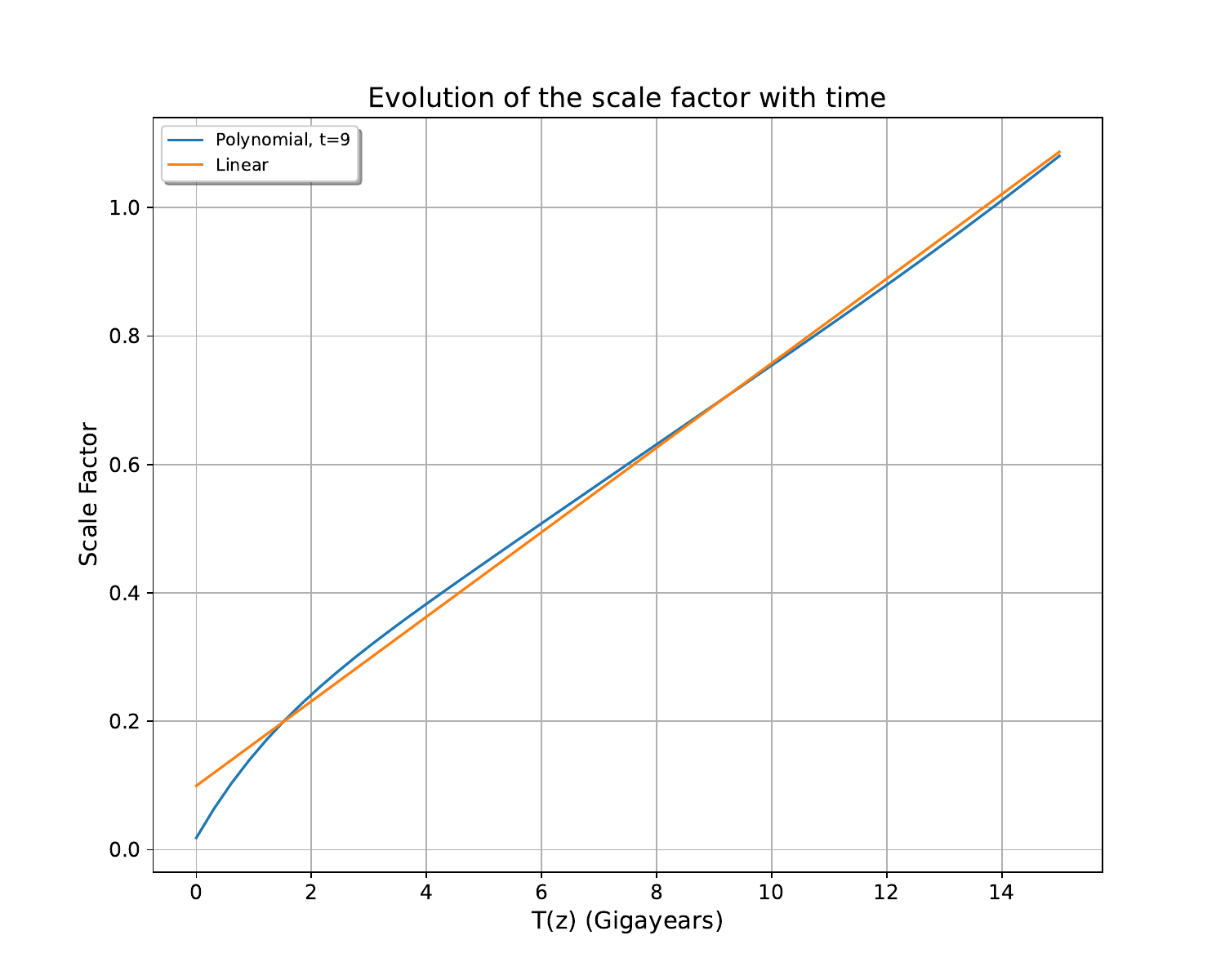}
\caption{Evolution of the scale factor $a(t)$ with time. The blue curve represents the best fit using a 9th-degree polynomial to reproduce its numerical evolution as precisely as possible, while the orange curve is the fit with a  straight line.}
\label{Figure_3}
\end{figure}

Considering    Eq. (\ref{expansion}), it is easy to see that  the first order  expansion works at low redshifts but it should get worse at higher values of $z$.  Notably, from  Eq. (\ref{H0_Planck}), we recover $T_0$ at higher redshifts. Furthermore, we may note that the evolution of $a(t)$, depending on the given cosmological eras dominated by different densities, is quite similar to a linear behavior, as  shown in Fig. \ref{Figure_3}.  Here we compare the evolution of $a(t)$, obtained by a  9$^{th}$ degree polynomial fit,   with  a straight line. We may note that the two curves are remarkably similar for a large range of  $T(z)$ values and easily converge to  the above limit of $H_0^z$ at high redshifts.

\begin{figure}
\includegraphics[width=0.45\hsize,height=0.45\textwidth]{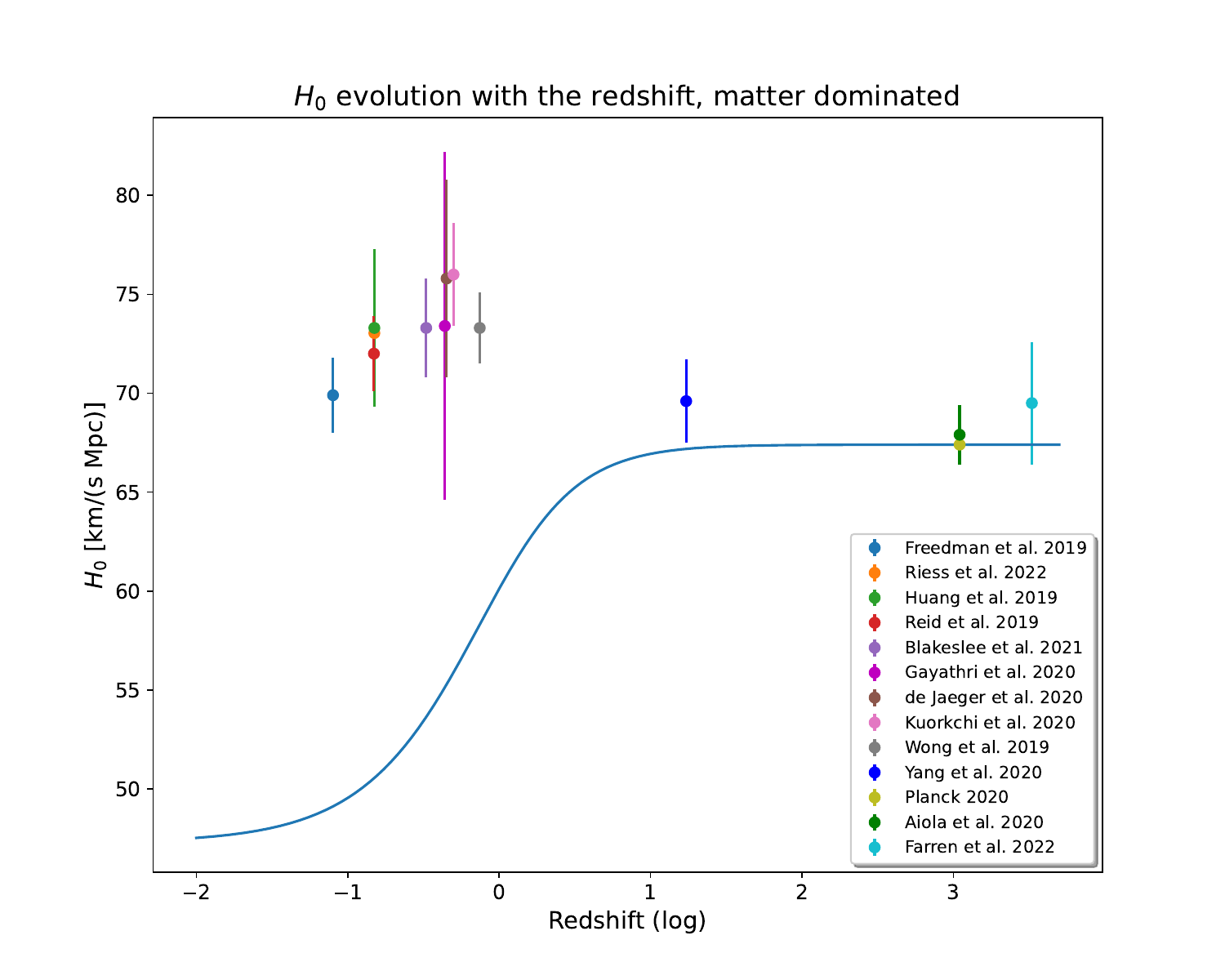}
\includegraphics[width=0.45\hsize,height=0.45\textwidth]{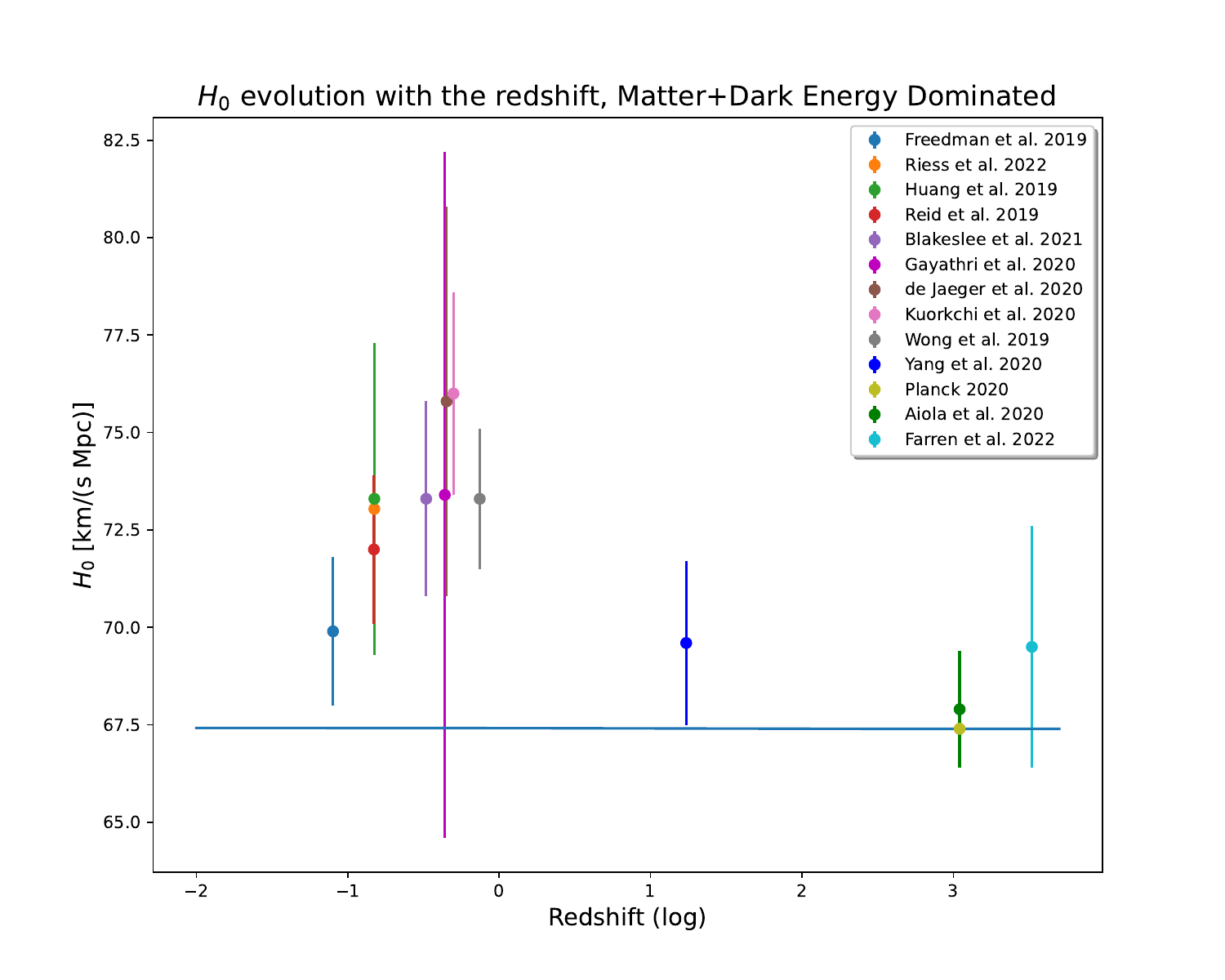}
\caption{Left panel: the value of $H_0$ derived  considering  $T(z)=\frac{T_0}{(1+z)^{3/2}}$  as a function of the redshift, confronted with observational data. Right Panel: the same comparison, but with a parameterization consistent with the Universe dominated by matter and dark energy. The $x$-axis is in logarithmic scale. We recall that the measurements have been taken from \cite{Riess_2022, Planck2020, Freedman_2019, de_Jaeger_2020, Huang_2020, Kourkchi_2020, Blakeslee_2021, Farren_2022, Reid_2019, Wong_2019, Aiola_2020, Yang_2019, Gayathri_2020}}
\label{Figure_4}
\end{figure}

To further validate our claims, we can try different parameterizations, which may be derived from different assumptions on the dominating cosmological densities in the function $E(z)$. This means that we do not consider our labeling, but specific approximations of the cosmological models. By taking into account the  definition on $T(z)$ in   Eq.\eqref{look-back_time}, via the time-evolution integral, one can derive

\begin{equation} \label{T(z)_integral}
T(z)=\frac{1}{H_0} \int_z^\infty \frac{dz'}{(1+z')E(z')}~.   
\end{equation}
In general, this integral has to be solved numerically, but it is possible to find simple analytical formulas linking it to $T_0$ for specific approximations of $E(z)$.
Let us start from a matter-dominated Universe. In  this case, it is $E(z)=\sqrt{\Omega_M(1+z)^3}$, from which $T(z)$ becomes
\begin{equation}
    T(z)=\frac{1}{H_0} \int_z^\infty \frac{dz'}{(1+z')^{5/2}}\,. 
\end{equation}
It can be solved analytically and one finds that the following parameterization is exactly valid \cite{sazhin2011}:
\begin{equation} \label{T(z)_matter_dominated}
    T(z)=\frac{T_0}{(1+z)^{3/2}}.
\end{equation}
If we introduce this new parameterization into our equations and compare the derived $H_0^{(z)}$ with the actual measurements, we obtain  results shown in the left panel of Fig. \ref{Figure_4}. It is worth noticing that this approach does not perform well in this case, as it produces a theoretical curve for $H_0$ that yields unreasonable results for low values of $z$, and it is not consistent with the late-type measurements, even for redshift regions where the Universe may be considered matter-dominated. 

If we, instead, consider a flat Universe with only dark energy and matter components (that is what we observe for the vast majority of the Universe lifetime,  neglecting the radiation contribution), it is 
$E(z)=\sqrt{\Omega_M(1+z)^3+\Omega_{\Lambda}}$.  We find, starting from the equations derived in \cite{Weinberg_1989} for this particular case, that $T(z)$ can be expressed as
\begin{equation}\label{T(z)_matter_dark_energy_formulation}
    T(z)=\frac{2}{3H_0}\biggl(1+\frac{\Omega_{M}}{\Omega_{\Lambda}}\biggr)^{1/2}\sinh^{-1}\biggl[\biggl(\frac{\Omega_{\Lambda}}{\Omega_{M}}\biggr)^{1/2}(1+z)^{-3/2}\biggr].
\end{equation}
From this equation, $T_0$ can be easily recovered in the limit $z \rightarrow 0$. In other words,   the relation between $T(z)$ and $T_0$ is as follows:
\begin{equation} \label{T(z)_matter_dark_energy}
    T(z)=\frac{\sinh^{-1} \biggl[\biggl(\frac{\Omega_{\Lambda}}{\Omega_{M}}\biggr)^{1/2}(1+z)^{-3/2}\biggr]}{\sinh^{-1} \biggl[\biggl(\frac{\Omega_{\Lambda}}{\Omega_M}\biggr)^{1/2}\biggr]}T_0.
\end{equation}
By introducing the last equation in our approach, we obtain the results shown in the right panel of Fig. (\ref{Figure_4}). We observe that the derived estimate for $H_0^z$ is independent of the redshift and aligns with the measurements provided by the Planck collaboration. However, it is not consistent with  measurements of $H_0$ obtained at lower redshifts. It is worth noticing that this estimate depends on the ratio $\Omega_{\Lambda}/\Omega_M$, which, in our case, is fixed to the values provided by Planck, but  can be modified for other inferred values of these quantities.
 
\begin{figure}
\includegraphics[width=0.45\hsize,height=0.5\textwidth]{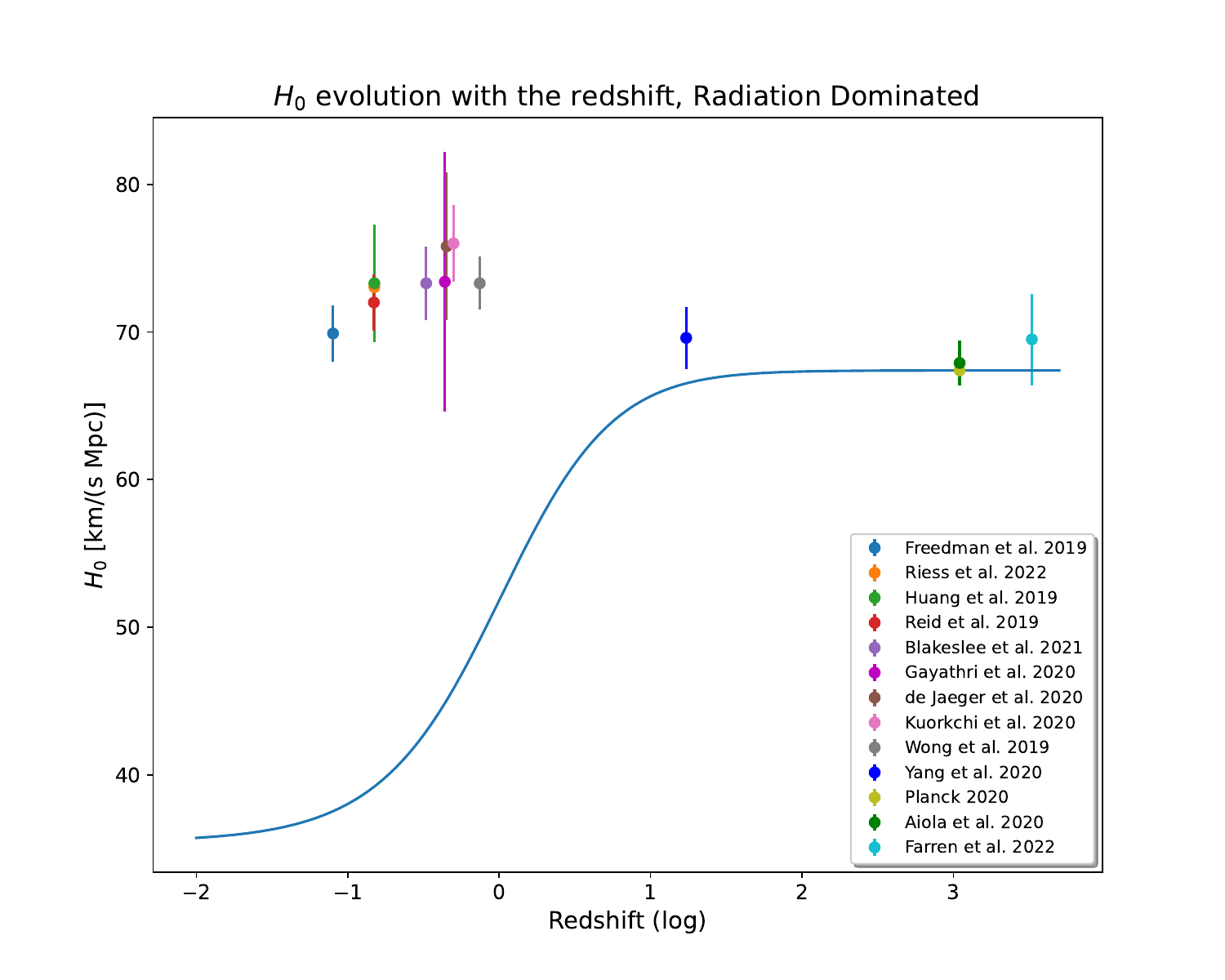}
\includegraphics[width=0.45\hsize,height=0.5\textwidth]{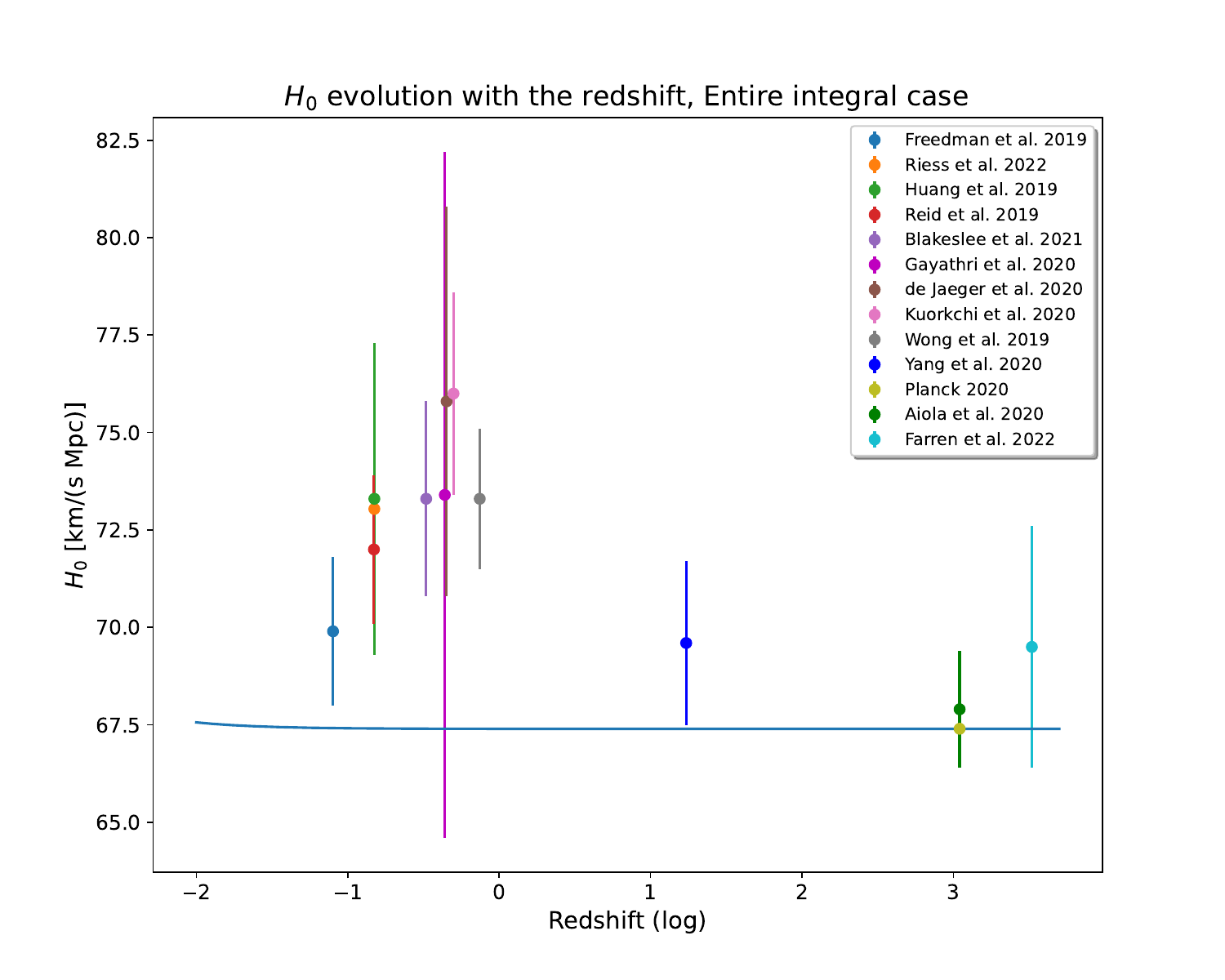}
\caption{Left panel: The value of $H_0$ derived from  $T(z)=\frac{T_0}{(1+z)^{2}}$,  plotted against the redshift and compared with observational data. Right panel: The same comparison, but taking into account the integral definitions of $T_0$ and $T(z)$. The $x$-axis is in logarithmic scale. We recall that the measurements have been taken from \cite{Riess_2022, Planck2020, Freedman_2019, de_Jaeger_2020, Huang_2020, Kourkchi_2020, Blakeslee_2021, Farren_2022, Reid_2019, Wong_2019, Aiola_2020, Yang_2019, Gayathri_2020}}
\label{Figure_5}
\end{figure}

As additional case, we have considered a radiation-dominated Universe, where 
$E(z)=\sqrt{\Omega_r(1+z)^4}$ and 
\begin{equation} \label{T(z)_radiation_dominated}
    T(z)=\frac{T_0}{(1+z)^{2}}\,.
\end{equation}
Furthermore,  without any assumption, we can  directly use the integral definitions of $T(z)$ and $T_0$. The plots of these two cases are shown in Fig. (\ref{Figure_5}). 

In the left panel, it is reported  the result obtained for the  radiation-dominated Universe. Notably, we observe unreasonable values for $H_0$  even with respect to the matter-dominated case. On the right panel, we show the result obtained adopting the general integral for $T(z)$, that is Eq. (\ref{T(z)_integral}).
Here, we obtain a constant value for $H_0$ around $67.4\, \text{km/(s Mpc)}$, independent of the redshift. This is because, by adopting the integral, we fall into a circularity problem which gives the $H_0$ provided by the Planck Collaboration at all redshifts. This result is also remarkably consistent with the one obtained by  Eq. (\ref{T(z)_matter_dark_energy}) for the Universe dominated by dark energy and matter. 

Interestingly, we emphasize that the parameterization presented in Eq. (\ref{T(z)}) is the only one consistent with both early and late-type measurements. In this framework, it can be exactly derived from an empty Universe with $E(z)=\sqrt{\Omega_k(1+z)^2}$. However, it is essential to note that such a Universe is in severe disagreement with cosmological observations \cite{Planck2020}. This discrepancy highlights the  conceptual difference between our labeling approach and the derivation of $T(z)$ through the integral in Eq. (\ref{T(z)_integral}).

In summary, the parameterization shown in Eq. (\ref{T(z)}) has been chosen as a  labeling process connecting the age of the Universe at various epochs through a point-by-point approach. Mathematically, it can be derived from a reliable approximation of the scale factor as a function of time, working also for high-redshift. Among the different tests, it stands out as the only parameterization that aligns with the $H_0$ observations at both early and late epochs. It is worth stressing  that, apart  our   labelling  $T(z)$,  the other tests have been performed considering  other parameterizations. This does not exclude the possibility of choosing other relations between $T(z)$ and $T_0$ which involve more complex functions  as $T(z)=T_0/(1+P(z))$ where $P(z)$ can be some function  of the redshift (e.g. a polynomial)  as considered in  cosmography \cite{Weinberg:1972kfs}. From  this more general approach, one could obtain  realistic models capable of  matching better the cosmic history.

\subsection{A variable \texorpdfstring{$H_0$}{}  from late-type estimates}

\begin{figure}
\includegraphics[width=0.45\hsize,height=0.5\textwidth]{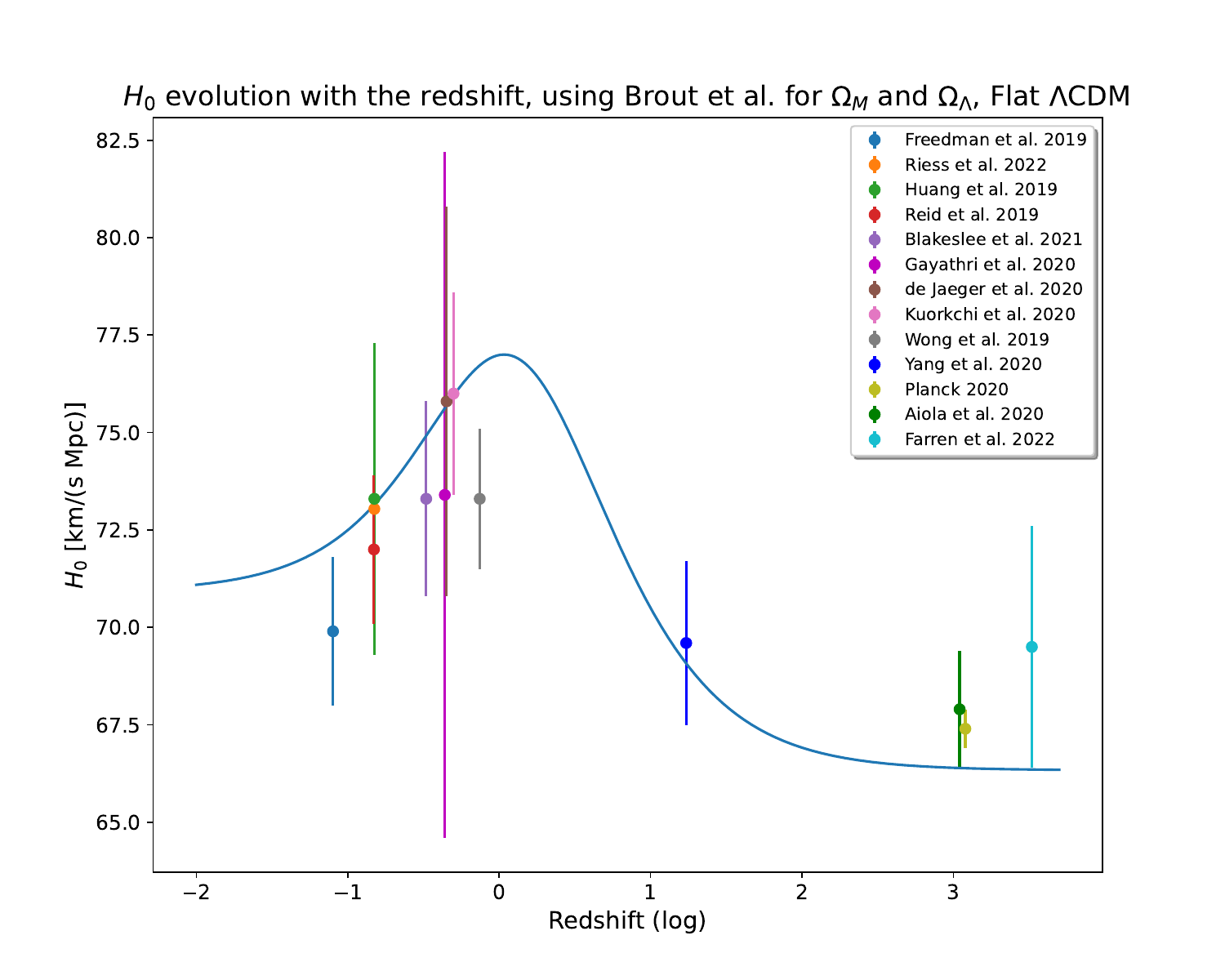}
\includegraphics[width=0.45\hsize,height=0.5\textwidth]{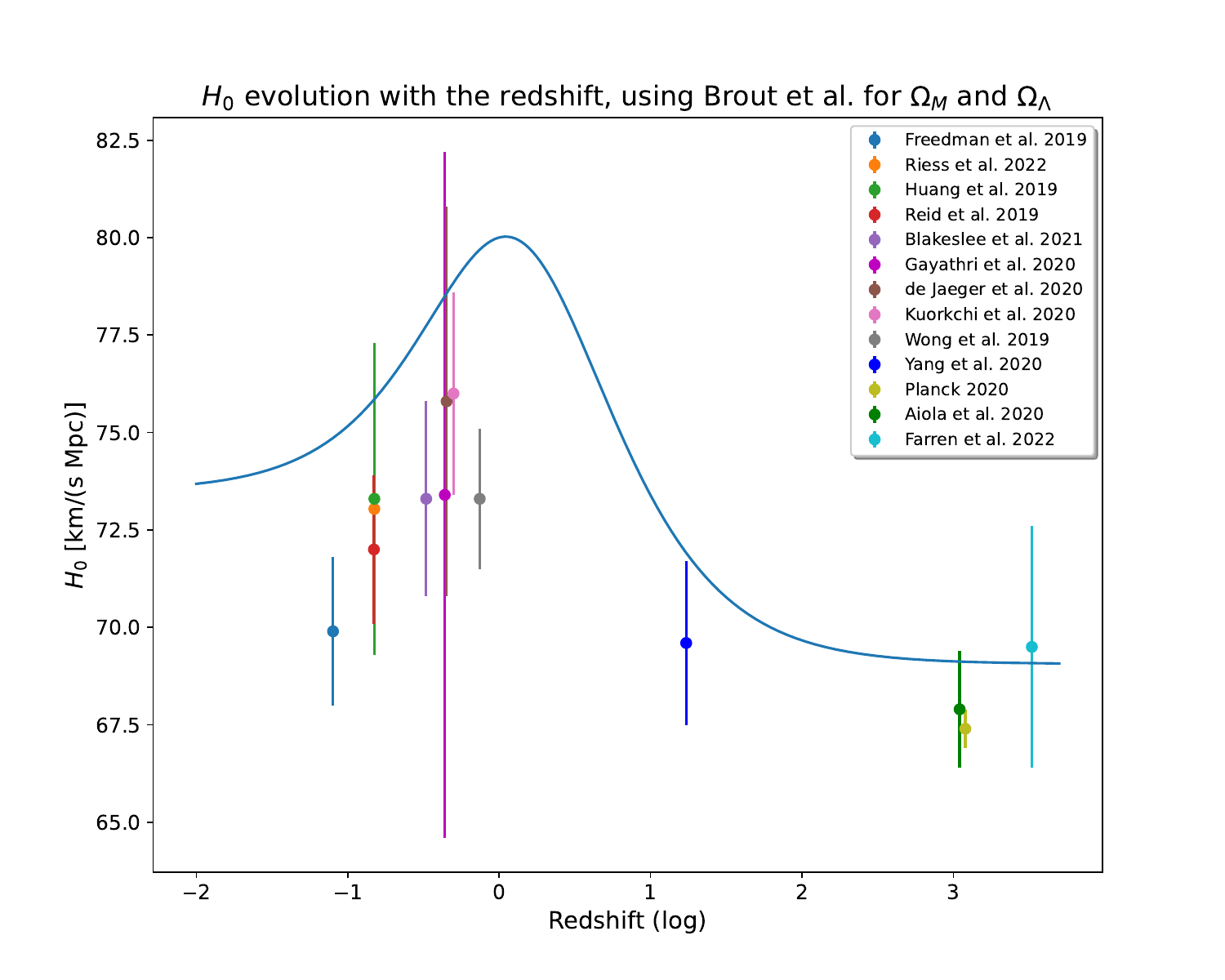}
\caption{Left panel: the value of $H_0$ derived from our methodology considering  $T(z)=\frac{T_0}{(1+z)}$  plotted against the redshift and compared with observational data, starting from the Pantheon+ results for a flat $\Lambda$CDM model. Right panel: the same comparison, but relaxing the flatness assumption. The $x$-axis is in logarithmic scale. We recall that the measurements have been taken from \cite{Riess_2022, Planck2020, Freedman_2019, de_Jaeger_2020, Huang_2020, Kourkchi_2020, Blakeslee_2021, Farren_2022, Reid_2019, Wong_2019, Aiola_2020, Yang_2019, Gayathri_2020}}
\label{Figure_6}
\end{figure}

The look-back time parameterization  of $H_0$ can be compared with late-type measurements for the cosmological parameters $\Omega_M$ and $\Omega_{\Lambda}$.
For this comparison, we will use  results obtained by the cosmological ladder approach involving the SNe Ia of the Pantheon+ set \cite{Scolnic_2022, Brout_2022}. In particular, we consider two different sets of results regarding $\Omega_M$ and $\Omega_{\Lambda}$.
In both cases, we  use the value for $T_0$ provided by the Planck results, since an estimate for this quantity cannot be derived directly from the cosmological computations reported in Ref. \cite{Brout_2022}. 
For the first set, we start from the values obtained in \cite{Brout_2022} for a flat $\Lambda$CDM model where they imposed $\Omega_M+\Omega_{\Lambda}=1$.   They found
\begin{equation}
      \Omega_M=0.334 \pm 0.018\,, \quad \Omega_{\Lambda}=0.666 \pm 0.018\,.
\end{equation}
Considering the best-fit values of these quantities in our model, and comparing the results with the aforementioned $H_0$ measurements, we find the results displayed in the left panel of Fig. \ref{Figure_6}.
We note how, in general, the curve obtained by our model is consistent with the majority of the $H_0$ measurements. However, we notice an interesting tension with the measurement provided by the Planck collaboration, which is, instead, remarkably matched by the curve displayed in Fig. \ref{Figure_2}.

Indeed, we note how, at high redshift, the value for $H_0$ provided by this model decreases to $66.34 \, \text{km/(s Mpc)}$, which is more than $2 \sigma$ in tension with the Planck measurement. An interesting observation is that this value is smaller than the measured one, contrary to what would be expected, given that late-type measurements usually translate with  higher estimates for $H_0$.

From this, we may conclude that the different values provided for $\Omega_M$ and $\Omega_{\Lambda}$ from early and late-type measurements have a significant impact on the comparison of our model with  measurements. 

This conclusion is further supported  by the second case we have considered, in which we have taken into account the results for $\Omega_M$ and $\Omega_{\Lambda}$ obtained in \cite{Brout_2022} relaxing the flatness assumption for the $\Lambda$CDM, for which
\begin{equation}
      \Omega_M=0.306 \pm 0.057\,, \quad \Omega_{\Lambda}=0.625 \pm 0.084\,,
\end{equation}
while for the value of $\Omega_r$, we rely on the value provided by Planck. Using our model for these starting values, we derive the right panel of Fig. \ref{Figure_6}. We observe a shift towards higher values of $H_0$ with respect to the results obtained in the left panel of the same figure and note that the curve is inconsistent with many observations, particularly those concerning the late Universe. In this comparison, it is important to point out the large uncertainties on $\Omega_M$ and $\Omega_{\Lambda}$ values, which could play a significant role.

\section{A variable \texorpdfstring{$H_0$}{}  in   \texorpdfstring{$\Lambda$CDM}{} model}

Let us discuss now the implications of our approach in the context of  $\Lambda$CDM model. As previously mentioned, a variable $H_0$  can be interpreted as a possible hint for the  breakdown of  FLRW metric and the $\Lambda$CDM model \cite{Krishnan_2021, Krishnan_2021b}.  Such a variation  can be explained in the context  of  ETGs \cite{CapozzielloETG,dainotti_2021a}. 
However, our analysis reveals that a variable $H_0$ can be entirely derived within the framework of  $\Lambda$CDM model, starting from some fundamental concepts  and from the parameterization  in Eq. (\ref{T(z)}), without making any a priori assumption on the evolution of $H_0$. This finding does not contradict the  role of $H_0$  in the Friedman equations because  the variation of this constant is  strictly connected to the redshift at which it has been measured. 

A robust confirmation of such a hypothesis would be attainable by conducting independent measurements  within the intermediate redshift range, where we observe the peak in our $H_0$ estimate as shown in Figure \ref{Figure_2}.

Notably, the Pantheon and Pantheon+ SNe Ia datasets \cite{scolnic-2018, Scolnic_2022}, spanning up to $z=2.26$, encompass a broad redshift range, albeit with a majority of SNe Ia situated in the low-redshift region. $H_0$ estimates already exist in this range, (e.g., \cite{Dainotti_2022e, Dainotti_2022f, Bargiacchi_2022}) but depend on calibration processes involving probes at lower redshifts. Consequently, promising new insights are anticipated from forthcoming investigations utilizing novel probes, such as quasars, gravitational wave standard sirens, galaxy clusters, Lyman-$\alpha$ lines, or GRBs (as reported in \cite{Dainotti_2022c, Califano1, Califano2, Bargiacchi_3, Bargiacchi_2023, Abbott_2017}).

In this context, the Euclid Mission \cite{Laureijs_2011, Scaramella_2022} could play a pivotal role. While its main focus is investigating the nature of dark energy via a wide set of observations of galaxy clusters and weak lensing phenomena \cite{Blanchard_2020} by also testing possible modifications of GR \cite{Frusciante_2023, Casas_2023}, it also holds the potential to provide precise estimates of cosmological parameters like $H_0$ within the intermediate redshift range. Additionally, given the wide survey expected  by Euclid, even though not primarily centered on transient phenomena such as SNe Ia, is likely to contribute significantly also to these types of observations \cite{Bailey_2023}.

\begin{figure}
\includegraphics[width=0.45\hsize,height=0.5\textwidth]{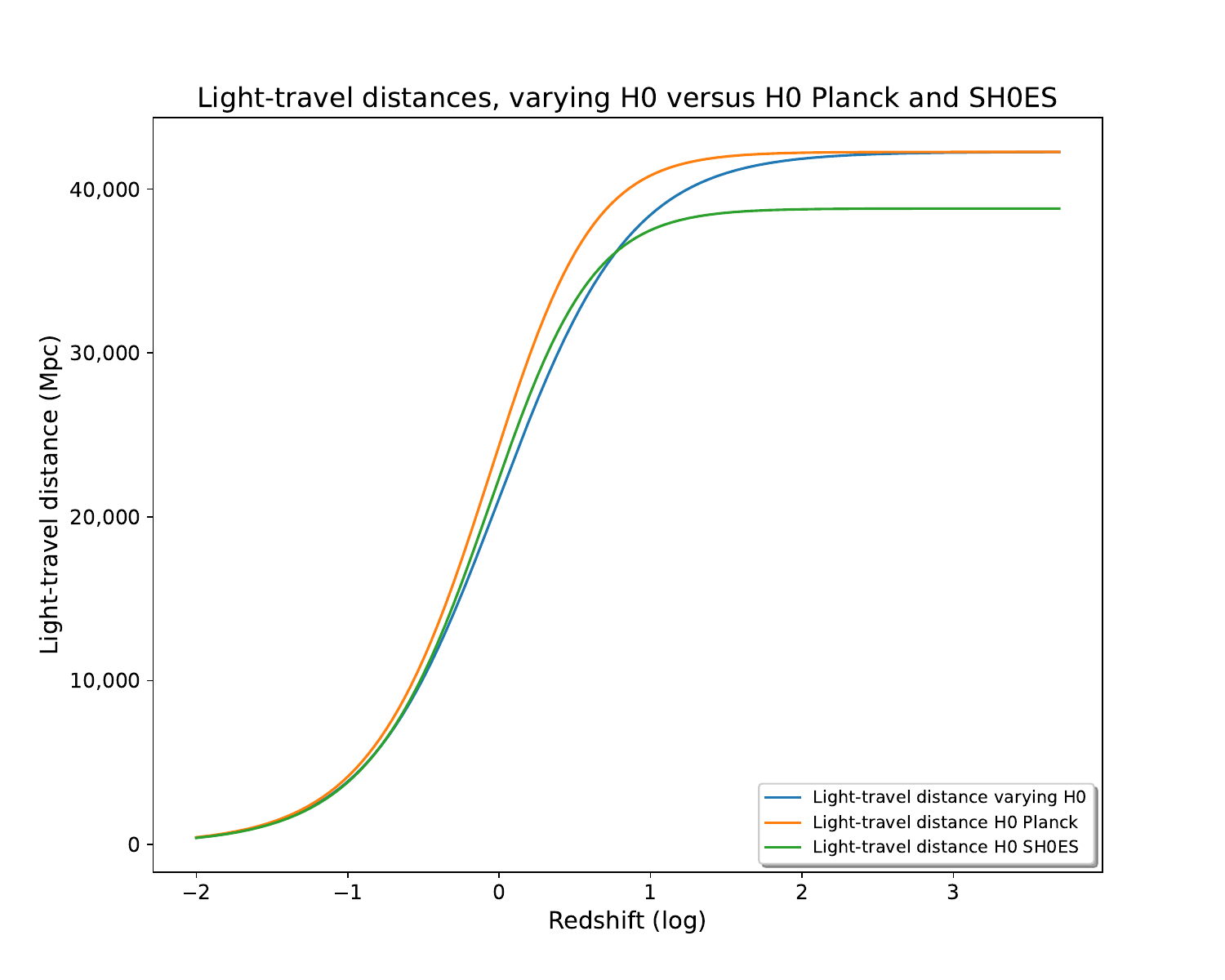}
\includegraphics[width=0.45\hsize,height=0.5\textwidth]{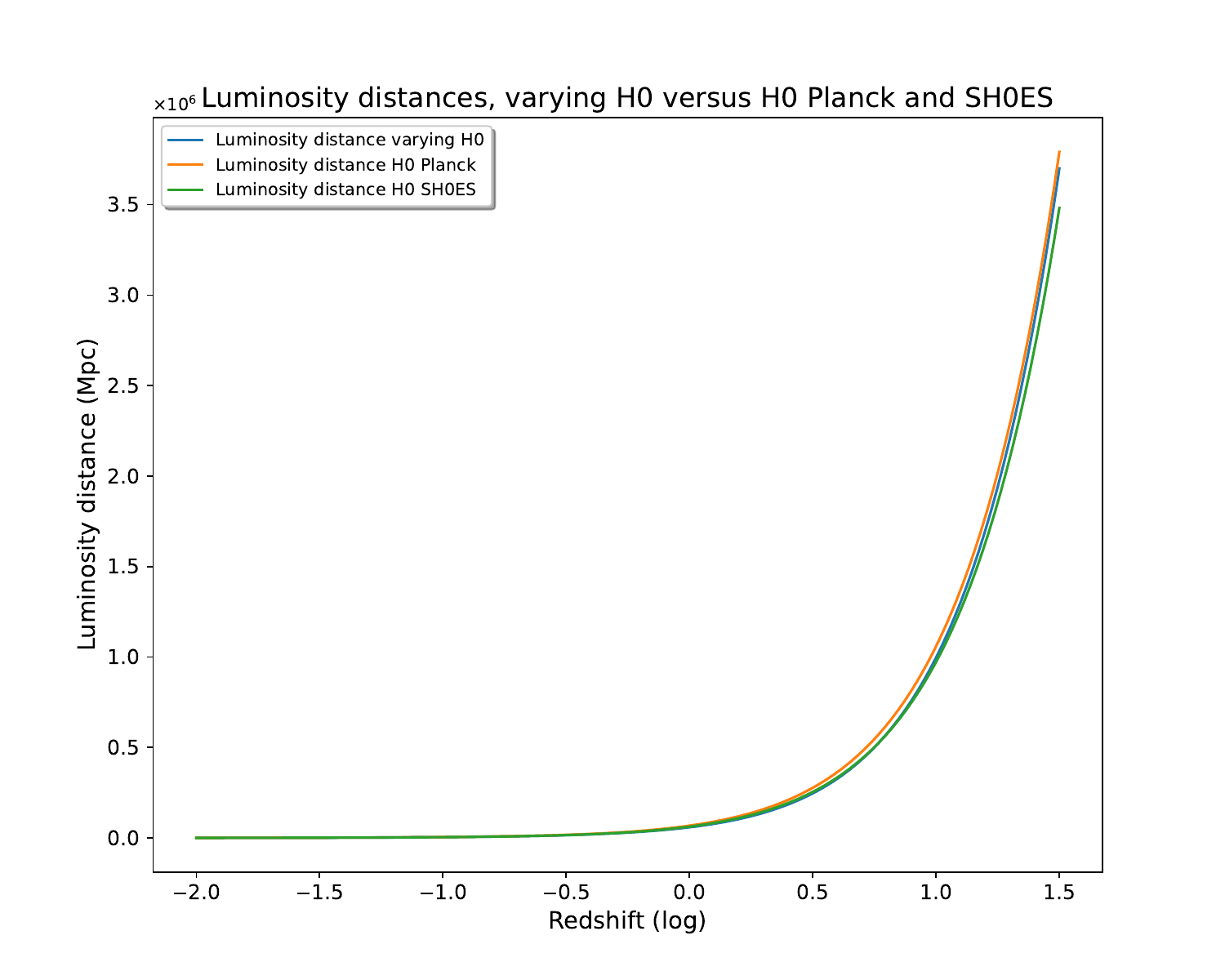}
\caption{Left panel: a comparison between the light-travel distance computed using the $H_0^{z}$ function (Eq. (\ref{H0_look-back})), with the distances derived by using the fixed values for $H_0$ provided by the SH0ES and Planck collaborations. Right panel: the same comparison as in the left panel, but  considering the luminosity distance. The $x$-axis is in logarithmic scale for both panels. Note that there are different scales on the x-axis in the two plots.}
\label{Figure_7}
\end{figure}

Let us discuss now the implications of a variable $H_0$  on the cosmological observations. To achieve this, we have investigated, as an example, how luminosity and light-travel distances behave with the $H_0^{z}$ function in Eq. (\ref{H0_look-back}), and compared these distances with the results obtained  by assuming the fixed values measured by the SH0ES and Planck collaborations. We selected the first distance because    it is arguably the most relevant for cosmological measurements, and the second due to its  connection with the look-back time. It is worth noticing that similar considerations apply to other cosmological distances. The results are presented in Figure (\ref{Figure_7}).

Both panels in this figure demonstrate how our approach naturally connects the two distances derived from fixed values of $H_0$ without significantly changing the overall behavior in the investigated redshift range. The effect is particularly evident in the left panel, where the light-travel distance is represented. We see how the distance computed by our approach is similar to the one derived from the SH0ES observations,  converging towards the Planck results at higher redshifts. This can be attributed to the role of $H_0$, which  is a normalization constant, not linked to any  astrophysical source. In conclusion, while a variable $H_0$ does impact the definitions of cosmological distances, it neither substantially changes the conclusions drawn from our method, especially at lower redshifts, nor  it changes the value of the distances so that it could be in contrast with  observations.

\section{Discussion and Conclusions}
In this work, we  provided, without claiming for completeness,  a summary of  $H_0$ measurements in both early and late-type frameworks, highlighting the existence of a tension independent of the particular probe or methodology used.  This independency emphasizes the decreasing probability that the tension stems from potential instrument-related biases or unaddressed systematic effects.

We explored how this tension has been tackled in the context of new physics, beyond the $\Lambda$CDM model. Various approaches have been proposed, yielding promising results. However, the main problem  is the substantial degeneracy among the different proposed frameworks, which does not allow us, up to now, to define a unique and definitive extension of the $\Lambda$CDM model. 
From this point of view, it is crucial to understand that, whichever the extended framework may be, it has to be able to reproduce  the $\Lambda$CDM results where the concordance among  the vast majority of measurements is  evident.
This check is essential in view to recover self-consistent models in agreement with cosmic history. 

A recent approach has brought significant attention to the possibility of a "running $H_0$" \cite{Krishnan_2021}, both in the observations context and in the interpretation of a quantity that is traditionally regarded as "constant". 

In this framework, we reported the analysis performed in \cite{Capozziello_2023}, which defines the Hubble constant via the look-back time.  It is possible to provide a general formula consistent with the measurements of this quantity both at early and late epochs just considering the related look-back time measurement of $H_0$ at any epoch.
Here, we delve further into this direction starting from  Eq. (\ref{T(z)}) and discussing the  consequences on the various cosmological distances. 

Specifically, we highlight that the formula $T(z) = T_0/(1+z)$ arises naturally through a point-by-point labeling of the age of the Universe at different redshifts, without the need to account for the cosmological evolution already incorporated into the function $E(z)$.

To validate our assumption, we show how it can be mathematically derived from a reliable approximation of the scale factor $a(t)$, and how other parameterizations, which can be derived from specific assumptions on the densities at various epochs, do not effectively fit observational measurements of $H_0$.

We have also considered different values of $\Omega_M$ and $\Omega_{\Lambda}$, taken by late-type results  from the SH0ES+Pantheon+ sets, finding significant effects in our comparisons, particularly regarding the value for $H_0$ provided by the Planck collaboration, which is not recovered in these cases. This observation is interesting because it allows us to conclude that the different measurements of $\Omega_M$ and $\Omega_{\Lambda}$, for  late and early epochs, introduce a notable tension. 

In particular, we derived the light-travel and luminosity distances as functions of our variable $H_0$ and compared the results with those obtained using the fixed values provided by early and late-time measurements. We found that, by our new definition, it is possible to  obtain distances able to link the SH0ES and Planck distances, without significantly modifying their overall behavior. In our opinion, this is an important check because the absence of unreasonable results, starting  from different distance definitions,  confirms the reliability of the approach.

An issue that may arise pertains to the measurements of the aforementioned $H(z)$ function and the extrapolation to the  $H_0$ value. In fact, one should be able to discern between variations related to the functional form of $H(z)$ and the variable nature of  $H_0$  at different redshifts.

An important point that have to be  emphasized  is that our results have been achieved entirely within the framework of the $\Lambda$CDM scenario, without requiring modifications or extensions. Theoretically, a variable $H_0$  can be interpreted as a breakdown of the FLRW metric. In our case,   $H_0$ is an integration constant related to the size of the Universe at a given redshift. In other words,  the  value of $H_0$ could  depend on the redshift at which it is measured, thus not undermining its  role in the cosmological equations but removing the tension issue. It is worth noticing  how a better  fit with respect to the  measurements can be obtained  by starting from the early estimate of  densities rather than the late ones.
However, this does not exclude the necessity of extending GR and  $\Lambda$CDM model, given  other issues  like the nature of dark energy and dark matter, and the lack of a self-consistent Quantum Gravity theory \cite{CapozzielloETG}. 

Observationally,  $H_0$ tension is not the only  tension in cosmology \cite{Abdalla:2022yfr}. For example, there is the so-called $S_8=\sigma_8 \sqrt{\Omega_M/0.3}$ tension, where $S_8$ is a parameter indicating the strength with which matter is clustered in the Universe. On this parameter, a discrepancy at $2-3 \sigma$ level \cite{Abdalla:2022yfr} exists between the measurements inferred by the Planck data and low-redshift probes, such as the weak gravitational lens and clusters of galaxies \cite{Joudaki_2016, Abbott_2022}. It is essential to investigate whether similar considerations to those discussed above can also apply to $S_8$ or if it represents an independent signal of deviations from the cosmological Standard  Model.

Furthermore, other measurements challenge the $\Lambda$CDM model, like some evidence of a possible non-zero curvature of the Universe \cite{Divalentino2020}, as well as anomalies in CMB observations, i. e., apparent correlations between  the Solar System plane and certain aspects of CMB. This evidence seems to provide  a preferred reference position  which should not be possible if the  Cosmological Principle is always valid  \cite{Challinor_2012, Schwarz_2016, Akrami_2020}.

In conclusion, our analysis suggests that new physics, in the form of extensions or modifications to  $\Lambda$CDM model, may not be necessary to address the specific issue of  $H_0$ tension, but it does not exclude its necessity for other fundamental issues. In a forthcoming study, we will discuss the other tensions under the standard of look-back time approach.

\section*{Aknowledgements}
This paper is based upon work from  the COST Action CA21136, ''Addressing observational tensions in cosmology with systematics and fundamental physics'' (CosmoVerse) supported by COST (European Cooperation in Science and Technology). The authors acknowledge the support by Istituto Nazionale di Fisica Nucleare, Sez. di Napoli, Iniziativa Specifica QGSKY. The authors  express their gratitude for  valuable  and constructive discussions  to Giada Bargiacchi, Micol Benetti,  Eoin Colgain, Maria Giovanna Dainotti, David Kraljic, Stefan Ruester,   Shahin Sheikh-Jabbari, and Alessandro D.A.M. Spallicci. Finally, we want  to thank the anonymous referees for their feedback and useful suggestions.

\bibliography{main_revised}

\end{document}